%% file: break171203.tex
\documentclass[epj]{svjour}
\usepackage[utf8]{inputenc}
\usepackage{epsfig}
\usepackage{amsmath}
\usepackage{braket}
\usepackage{enumerate}
\usepackage{hyperref}
 \usepackage{cleveref}
 \usepackage{cite}

\begin{document}
 
\title{Breaking of axial symmetry in excited heavy nuclei as identified \\
in Giant Dipole Resonance data}

\author{
E. Grosse\inst{1} \thanks{electronic address: e.grosse@tu-dresden.de}
\and  
A.R. Junghans \inst{2} \thanks{electronic address: a.junghans@hzdr.de}
\and
R. Massarczyk\inst{1} \thanks{\emph{present address:} Los Alamos National Laboratory, Los Alamos, New Mexico 87545, USA}
}

\institute{Institute of Nuclear and Particle Physics, Technische Universit\"at Dresden,
01062 Dresden, Germany \and Institute of Radiation Physics, 	
Helmholtz-Zentrum Dresden-Rossendorf, 01314 Dresden, Germany }

\abstract{
A recent theoretical prediction of a breaking of axial symmetry in quasi all heavy nuclei is confronted to a new critical analysis of photon strength functions of nuclei in the valley of stability. For the photon strength in the isovector giant dipole resonance (IVGDR) regime a parameterization of GDR shapes by the sum of three Lorentzians (TLO) is extrapolated to energies below and above the IVGDR. The impact of non-GDR modes adding to the low energy slope of photon strength is discussed including recent data on photon scattering and other radiative processes. These are shown to be concentrated in energy regions where various model calculations predict intermediate collective strength; thus they are obviously separate from the IVGDR tail. The triple Lorentzian (TLO) ansatz for giant dipole resonances is normalized in accordance to the dipole sum rule. The nuclear droplet model with surface dissipation accounts well for positions and widths without local, nuclide specific, parameters. Very few and only global parameters are needed when a breaking of axial symmetry already in the valley of stability is admitted and hence a reliable prediction for electric dipole strength functions also outside of it is expected. }
\maketitle

\section{Introduction}
\label{sec1}
The ongoing discussion \cite{ok12,wr12,to13,ni14} about triaxial shapes in heavy nuclei, recently often studied off stability, may provoke the question, how well the widely used assumption about axial symmetry of most less exotic nuclei is founded on sufficiently sensitive experimental data. A nonaxial shape of heavy nuclei is a less stringent assumption than the often made assumption of axiality, which probably originates from atomic hyperfine structure observations \cite{sc35}, which are usually made on unpolarized samples and thus are insensitive to broken axiality. Based on formal logics we assume that a support for the non-axiality of many heavy nuclei as predicted recently by Hartree-Fock- Bogoljubov calculations by Delaroche et al. \cite{de10} has to rely on a falsification of the prejudice of an axial or even spherical shape for them. This does not exclude the possibility, that several properties as observed even in many nuclei may be reproduced well by theoretical models based on axial symmetry. We will present in this paper how breaking axial symmetry influences the splitting of the isovector nuclear electric dipole resonance (IVGDR). In text books this feature is often considered an indicator of nuclear deformation, and our detailed investigation aims for facts conflicting an assumption of axiality. Such features will be regarded for more than 20 nuclei (as examples) in a wide range of mass number A and ground state deformation. 

Here the fundamental origin \cite{ge54} of the sum rule for the electromagnetic strength in nuclei will be discussed, and a parameterization of the IVGDR as seen in photonuclear reactions with heavy nuclei will be presented. In contrast to previous work the electric dipole strength is derived from IVGDR data without assuming axial symmetry as it is usually made \cite{di88}for most heavy nuclei; instead theoretical \cite{de10} deformation and triaxiality parameters will be used. As was noted recently for nuclei with mass number $A >70$ \cite{ju08}, the apparent width of the IVGDR is an important parameter in the characterization of especially its low energy tail.  From various experimental data on photon emission and absorption an empirical parameterization of photon strength in this tail region is derived; here intermediate structure has been observed since long \cite{ax70,ba73}. We will critically regard predictions made in the past on the energy-dependence of the IVGDR width and its eventual relation to the deformation induced split. Various experimental information on minor strength in the tail region, nowadays often related to 'pygmy' and other modes \cite{kn06,he10,gr12,sa13}, will be discussed in view of multipolarity and parameterized phenomenologically. This done in view of the possible influence on predictions on neutron capture for which such minor strength as well as the extrapolation from the IVGDR energies are of relevance. In sections \ref{sec5} and \ref{sec6} energy dependent dipole strength functions resulting from the combination of IVGDR and minor strength will be presented for nuclei in a wide range of even and odd A. As the respective experimental studies can often only be performed for nuclei in or close to the valley of beta-stability, we aim for a small number of global parameters to increase the reliability of any extrapolation.

\section{Nuclear deformation and spectroscopic data }
\label{sec2}
The electromagnetic response of nuclei has played an important role for the exploration of their size and the departure of their shape from spherical symmetry was first indicated by a hyperfine splitting of atomic transitions due to the nuclear electromagnetic field \cite{sc35} seen by the electrons. Apparently this was considerably larger than the one expected from the extra protons in odd nuclei. This finding has led to an extension of the nuclear shell model (which is based on one harmonic oscillator frequency for each major shell) to the ‘Nilsson model’ \cite{ni55}: This model distinguishes oscillations which are oriented parallel and perpendicular to the symmetry axis. This is a limitation to the case of axial symmetry of the core and the potential it exerts on the extra nucleons and it was explicitely pointed out that the model was derived for strongly deformed nuclei \cite{ni55}. As long as the normalized difference $\delta = \frac {\Delta{}\omega}{\omega_{av}}$, proposed as a measure of the deformation of the core, was sufficiently large, this approximation has explained the observed moments and transition rates ; it also led to a reasonably good understanding of level sequences and spins in odd nuclei \cite{ni55, mo55}. The level structure of even nuclei as well as the probability observed for their absorption or emission of photons was also predicted semi-classically by an axial rigid rotor model \cite{al56, bo57, na65, bo75}, showing a dependence on the deformation parameter. One serious shortcoming of this model is the fact, that it only predicts one ‘collective’ $2^+$-state. Experimentally at least two $2^+$-levels with enhanced transitions to the ground state are observed in nearly all even nuclei. This has led to the assumption \cite{da65, bo75} of a semiclassical coupling of the collective rotation to a collective quadrupolar vibration around a (possibly non-axially) deformed core described by deformation parameters $\beta \cong \delta$ and $\gamma$, which characterizes the non-axiality.  A breaking of axial symmetry in some nuclei was concluded from the calculation of fission barrier heights, as pointed out already long ago \cite{la73, gi82}. A fit to experimental spectroscopic data has indicated triaxiality to be important also for several odd nuclei \cite{me74}, and detailed multiple Coulomb excitation studies found permanent triaxiality for various nuclei in the valley of stability \cite{cl86, ma90, wu96, sr06}. To better understand higher spin states the concepts of wobbling modes, chiral or parallel bands and tilted cranking were introduced \cite{fr01, be03}; their appearence in many nuclei stresses the need to investigate axial symmetry breaking. We are not aware of such studies covering heavy nuclei globally; apparently difficulties arise from the absence of a distinguished axis to project on and probabely this prohibits an extension of the Nilsson approach \cite{ni55} to non-axial nuclei. A more microscopic approach seems justified to study a possible extension of it to enclose nuclei with smaller $\beta$ and non-zero~$\gamma$. \\

A recent theoretical prediction \cite{de10} as tabulated for a large number of nuclei indicates broken axial symmetry for quasi all of them and especially significant for nuclei previously often regarded transitional between axial and spherical in shape. Even for near magic nuclei triaxiality appears, but due to their small quadrupole deformation its effect is marginal; in some cases the also calculated standard deviation does not include $\gamma = 0$. These predictions are based on a self-consistent microscopic calculation in a Hartree-Fock-Bogolyubov scheme using the Gogny D1S interaction, and the wave functions are constrained to the selected values of Z and N. The also carried out generator coordinate method and the projection on good angular momentum are especially important at low angular momentum and the calculations generate at least two ‘collective’ $2^+$-levels in nearly all nuclei. Long ago it was pointed out \cite{ha84}, that such a quantum mechanically proper variation after projection may shift the $\gamma$-oscillation centered at axiality to $\langle\gamma\rangle\neq 0$. The calculations only assume $R_\pi$-invariance, {\em i.e.} wave functions stay unchanged after a rotation by 180 degree; they “are free of parameters beyond those contained in the Gogny D1S interaction” (adjusted to the properties of nuclei with deformation and based on “a density-dependent HFB approximation. They describe simultaneously the gross properties depending on the average field as well as the effects of pairing correlations via the Bogolyubov field with the same force” \cite{de80}). The split of the giant dipole resonances \cite{mo59, bo75} is related to the oscillator parameters corresponding to the nuclear shape and its axiality. The similarity between observations and the CHFB-calculations, as shown in the corresponding paper \cite{de10}, and the spin projection led us to use these as a reference. As discussed below, the equivalent sphere radius $R_0=\sqrt{(5/3)} \langle R_p \rangle$, with the tabulated \cite{de10} point proton radius $R_p$, will enter into our predictions for the IVGDR as well as the tabulated $\beta$ and $\gamma$ values which are related to the three
oscillator frequencies (controlling the IVGDR splitting) by the following relations:   
\begin{align}
	\frac{\omega_x}{\omega_y} &= \exp\left(-x_b\sqrt{3}\sin{\gamma}\right) \label{eqPQ}\nonumber \\
	\frac{\omega_x}{\omega_z} &= \exp\left(x_b\left[\frac{3}{2}\cos\gamma-\frac{\sqrt{3}}{2}\sin\gamma\right]\right)\\ 
	{\rm with} \quad x_b &= \frac{\beta}{2\beta + 1} \quad {\rm and} \quad {\omega_0}^3 ={\omega_x}{\omega_y}{\omega_z}\nonumber 
\end{align}

The last line uses the concept of the conservation of the nuclear density in an equivalent ellipsoid to a sphere with the same volume $V= 4/3\pi R_0^3$ with radius parameters $R_k$, which are inversely proportional to the harmonic oscillator constants $\omega_k$. In the paper \cite{de10} formulae are given to derive the usual deformation parameters $\beta$ and $\gamma$, which are also listed with their variances in the attached supplemental material as given for 1712 even-even nuclei.  It could been shown numerically, that for the small $\beta$ in the range of interest the differences between the deformation parameters used here and the Hill-Wheeler \cite{hi53} values used by us before \cite{ju08} are below $3 \%$ and hence not significant; a similar result was found for two popular parameter choices \cite{ni55, bo75}, the second of which is not volume conserving. 

\section{Photon absorption by nuclei}
\label{sec3}
In addition to the direct Thomson scattering a photon of sufficiently high energy $E_\gamma$ excites nuclei from the ground state resonantly; this is described by a Lorentzian centered at the resonance at $E_r$ with total width $\Gamma_r$:		
\begin{equation}
\sigma_\gamma \cong I_{r0} \frac{2}{\pi}\frac{E_\gamma^2\Gamma_{r}}{(E_r^2-E_\gamma^2)^2 + E_\gamma^2\Gamma_{r}^2} 
\label{eqsig}
\end{equation}

The integral of the absorption cross section $\sigma_\gamma$ over the resonance with spin $J_r$ is denoted by $I_{r0}$:
\begin{equation}
I_{r0}=\int\sigma_\gamma(E_\gamma)dE_\gamma= \frac{g(\pi\hbar c)^2 \Gamma_{r0}}{E_r^2}; \quad g = \frac{2J_r+1}{2J_0+1}  
\label{eqIr0}
\end{equation}
where $\Gamma_{r0}$  is the partial width of the transition between the resonant level $(E_r ,J_r)$ and the nuclear ground state $(0,J_0)$. It is directly proportional to the square of the electromagnetic transition matrix element; a respective relation exists for electric and magnetic excitation with multipole order $\lambda=1$:	
\begin{equation}
\Gamma_{r0}(E_\gamma;{E,M}1) =  \frac{16\pi}{9} \frac{\alpha E_\gamma^3}{g (\hbar c)^2}
 |\braket{r\|{\bf E,M}1\|0}|^2 
\label{eqGr0}
\end{equation}
Derived from very general conditions as causality and analyticity together with dispersion relations from QED the interaction of short wavelength photons with nuclei of mass number $A = Z+N $ can be ‘integrated up to the meson threshold’ analytically, leading to the energy-weighted sum rule of Gell-Mann, Goldberger and Thirring (GGT) \cite{ge54}:
\begin{align}
  I_{\rm A} &= \int_0^{m_\pi c^2} \sigma_\gamma (E_\gamma) dE_\gamma\nonumber\\
 &\cong 2\pi^2 \frac{\alpha \hbar^2}{m_n}\left [\frac{ZN}{A}+\frac{A}{10}\right ]
 \label{eqIE1}\\
&\cong 5.97\left [\frac{ZN}{A}+\frac{A}{10}\right ]{\rm MeV~fm^2} \nonumber 
\end{align}
Here $m_n$ and $m_\pi$ stand for the mass of nucleon and pion, respectively and no arguments \cite{le50} about the nuclear absorption of photons with energies above $m_\pi c^2$ are needed. The second term “contains all of the mesonic effects” and is assumed \cite{we73} to be accurate within $30\%$. It was approximated by assuming “that a photon of extremely large energy interacts with the nucleus as a system of free nucleons”, and a correlation to hadronic shadowing was investigated to be weak \cite{we73}. Eq. (\ref{eqIE1}) includes all multipole modes of photon absorption and the first term in the sum is identical to the “classical (TRK) sum rule” for electric dipole radiation \cite{ku25, re25} as the contribution to other multipoles will be shown to be small.  Absorption by the nucleons does not contribute below $E_\gamma = m_\pi c^2$, but nucleon pairs and especially p-n-pairs are strongly dissociated by photons with $20<E_\gamma<200$ MeV. The respective “quasi-deuteron effect” has been derived from the expression valid for the free deuteron by correcting for Pauli blocking \cite{ch91}. 	\\

\begin{figure}[ht]
\includegraphics[width=1\columnwidth]{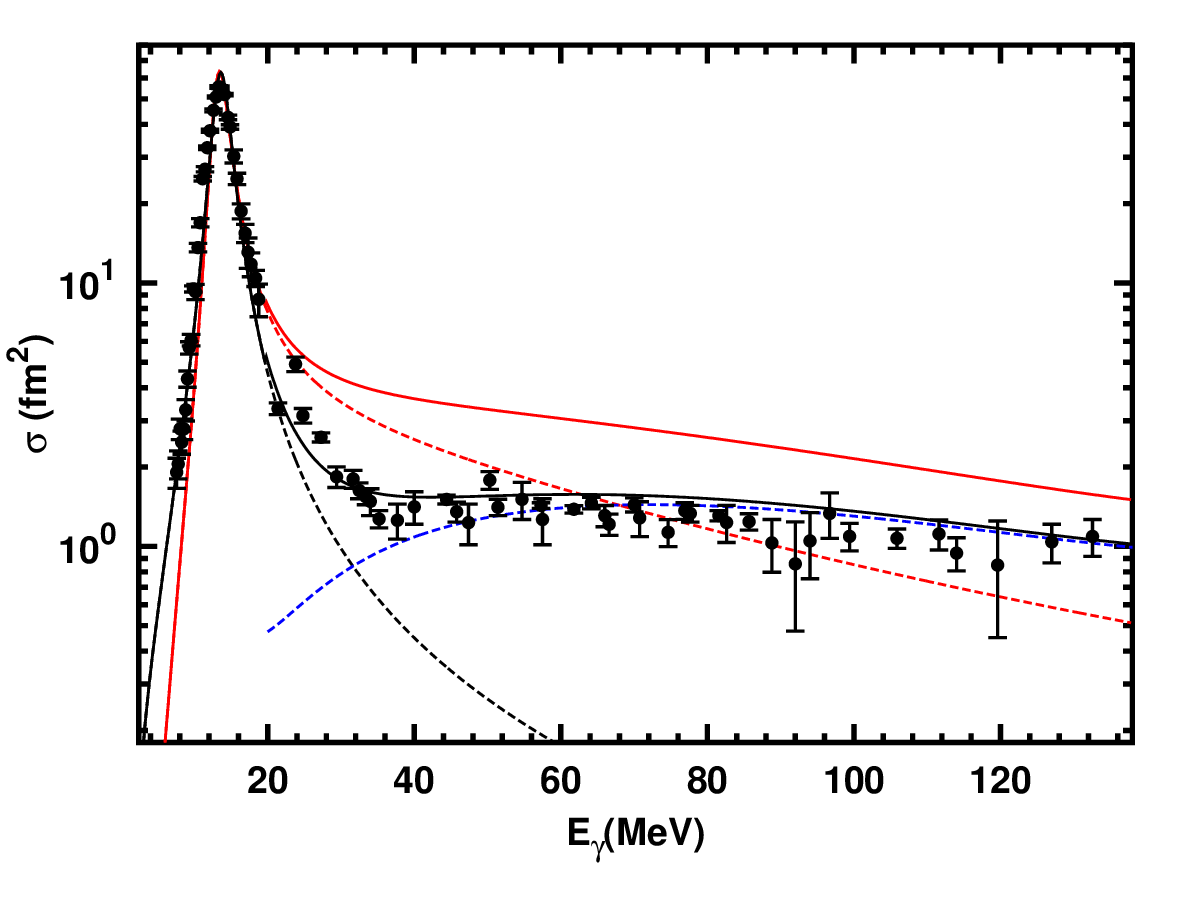}
\caption{(Color online)Cross section of photo-neutron production data \cite{ah85} on $^{208}$Pb in comparison to a Lorentzian for the isovector IVGDR (black and red lines, see text) and the quasi-deuteron effect (blue dashed line). The sum of both contributions is given as drawn lines. In $^{208}$Pb a deformation induced widening can be neglected as will become obvious in Fig.~\ref{figPb}. } 
\label{figPh}
 \end{figure}

The  photon absorption of nuclei is dominated by an excitation of a giant resonance, the IVGDR, which represents a strongly collective oscillation of neutrons against protons \cite{go48, st50}. Photo-disintegration is the main channel to observe the IVGDR and the good agreement to such data in the region below 20 MeV is seen in Fig.~\ref{figPh}. It indicates that the first term in Eq. (\ref{eqIE1}), a Lorentzian in accordance to the classical electric dipole sum rule (TRK, \cite{ku25, re25}) is a good ansatz to describe the IVGDR. Photo-neutron data are available \cite{ah85} for $^{208}$Pb up to energies above $m_\pi c^2$ as shown in Fig. \ref{figPh}. They are compared on an absolute scale to a Lorentzian like in Eq. (\ref{eqsig}) with pole energy $E_r = 13.6$ MeV and $I_r$ normalized such that it’s integral agrees to the first term in Eq. (\ref{eqIE1}). For $E_\gamma > 20 $MeV a comparison to the expression for the absorption cross section corresponding to the quasi-deuteron mode\cite{ch91} is shown, also on absolute scale.

The sum of both is also depicted and the case of a constant width $\Gamma_r  = 3.0$ MeV is indicated in black. In Fig. \ref{figPh} also the effect of a gamma energy dependent width is illustrated in red; this curve evolves from making the width proportional to the square of the photon energy $\Gamma_r  \sim (\frac{E_\gamma}{E_r})^2$. The latter proportionality was proposed \cite{ka83} to evolve from Landau theory of Fermi liquids. Obviously the data above 25 MeV are clearly below the curve corresponding to an IVGDR-Lorentzian with an increasing width. The cross section is well reproduced without such a change and its integral agrees to the sum of both terms in Eq. (\ref{eqIE1}); it is worth mentioning that a similar situation was discussed \cite{ch91} for other nuclei. The cross section observed to slightly exceed the sum between 20 and 30 MeV may be identified with the contribution from the IVGQR as shown in section~\ref{sec6}. In contrast to elastic electron scattering \cite{pi74, do82} the cross section depicted in Fig.~\ref{figPh} shows only weak signs of giant resonances of multipolarity other than E1. The disagreement especially in the high energy slope as seen in Fig.~\ref{figPh} is obvious when a proportionality between width and photon-energy squared is imposed; hence the KMF-method of treating the IVGDR width \cite{ka83} is falsified at least for $^{208}$Pb. \\

The KMF-model was allegedly derived from the theory for Fermi liquids although a detailed relation to the work of Landau or Migdal is not given in that work \cite{ka83}, which explicitly states below its Eq. (29), that a "direct comparison" between the "spreading width" in a Lorentzian for the GDR and the one of the theory of Fermi liquids is "difficult…and not clear" \cite{ka83}, and they further assume {\em ad hoc} without any additional arguments, that they coincide. The KMF-model was favoured within the Reference Input Parameter Library (RIPL) project \cite{ca09} to improve the agreement to some data below the IVGDR. But at variance to the original paper \cite{ca09}, a more recent work \cite{pl11} from that collaboration now proposes to insert a linear decrease of  $\Gamma_r$ with ${E_\gamma}$ into Eq. (\ref{eqsig}) instead of a quadratic one. Also work published some time ago was not in favour of the KMF-model: Fundamental theoretical arguments have been used to show, that "Landau damping is not the appropriate process for describing the damping of the low-multipole giant resonances" \cite{fi86}. It was demonstrated \cite{be95} for the nucleus $^{163}$Dy that the KMF-model does not work, when 2-step cascade data are analyzed using a double Lorentzian for the IVGDR; also for $^{157}$Gd (see \cref{figGd}) this model was not favoured by the authors of ref.(\cite{ko93}, who had first proposed its use earlier. \\

Although the IVGQR is not a distinct quantum level, but a sum of densely packed levels resonantly enhanced, it can be described by a sum of Lorentzians \cite{da64}. The possibility of a Lorentzian was tested numerically at hand of data for a few nuclei \cite{go77}. Derived from Eq. (\ref{eqsig}), but now applied to a wide giant ‘collective mode’ forming an envelope over narrow electric dipole states excited by E1 radiation, Eq. (\ref{eqdIE}) will be used for the parameterization of the IVGDR, with $k$ characterizing a deformation induced split. For the sum of $k$ Lorentzians, the main term of Eq. (\ref{eqIE1}), the classical sum rule for E1, is divided equally into $k$ fractions, assuring a normalization of the integrated strength:	
 
\begin{align}
  \frac{dI_{\rm E1}}{dE_\gamma} (E_\gamma) 
	&\equiv \sigma_{abs}^{{\rm E1,IV}}(E_\gamma) \label{eqdIE}\\  
&\cong 5.97\frac{ZN}{A}\frac{2}{k\pi} 
\sum_{i=1,k} \frac{E_\gamma^2\Gamma_i}{(E_i^2-E_\gamma^2)^2 + E_\gamma^2\Gamma_i^2} 
{\rm fm}^2 \nonumber
\end{align}
 
In Fig.~\ref{figPh}, asssuming $^{208}$Pb to be spherical even above 10 MeV, $k =1$ was used. It will be shown in section \ref{sec4}~and~section \ref{sec6}, that the TRK sum rule can be well fulfilled for all heavy nuclei, when account is made for a universal breaking of axial symmetry leading to $k =3$. As long as the sum rule is respected, the extraction of dipole strength in the region of the maximum and of the height of the low energy tail are fixed unambigously by the use of radii, deformation and triaxiality from the CHFB calculations \cite{de10}. Using Eq. (\ref{eqEi}), the energies of the three resonance poles are derived from the spherical centroid energy  $E_0=E_{IVGDR} $ and the well known \cite{bo57} proportionality between $E_i$ and $\omega_i$. For $E_0$ of Eq. (\ref{eqEi}) we note that two historic theoretical treatments of the IVGDR predict its energy rather well for medium mass nuclei \cite{go48}, respectively for the very heavy ones \cite{st50}. By using concepts of the droplet model these two approaches were unified \cite{my77}. The symmetry energy $J=32.7$ MeV and surface stiffness $Q=29.2$ MeV are taken from the finite range droplet model \cite{mo06, mo08} and the IVGDR centroid energies $E_0(Z,A)$ will be shown for $78 < A < 254$ to be well predicted with only one additional quantity, an effective nucleon mass. It was adjusted in an overall fit to the IVGDR positions and we obtained  $m_{eff} = 800 {\rm MeV}/c^2$, which differs from our earlier work \cite{ju08} where $874{\rm MeV}/c^2$ was used. This change is due to the different choice of the nuclear radius as $R_0=\sqrt{(5/3)}\cdot\langle R_p\rangle$ with the point proton radius $\langle R_p\rangle$ taken from the CHFB calculations \cite{de10}.  Hence only very few parameters for the centroid energy of the IVGDR's are required in a global desccription. As previously \cite{ju08, ju10} we follow \cite{my77} and use (with units MeV and fm):  
 
\begin{align}
E_0 &= \frac{\hbar c}{R_0} \sqrt{\frac{8 J}{m_{eff}} \cdot \frac{A^2}{4 N  Z}} \left[1 + u - \varepsilon \cdot \frac{1 +\varepsilon + 3 u}{1 +\varepsilon + u}\right]^{-1/2}\nonumber \\
\varepsilon &=  0.0768,  \quad u = \left(1 - \varepsilon \right) \cdot A^{-1/3} \cdot \frac{3 J}{Q}\label{eqEi} \\ 
E_i &= \frac{\omega_i}{\omega_0}\cdot E_0  \quad {\rm and } \quad\Gamma_i = c_w E_i^{1.6}\nonumber
\end{align}

The nature of the IVGDR does not allow for the direct determination of its Lorentz widths $\Gamma_i$ in analogy to Eq. (\ref{eqGr0}), but it was predicted theoretically \cite{do72} to be related to nucleon dissipation in nuclei. Hydro-dynamical considerations \cite{bu91} predict the dependence of the damping width $\Gamma_i$ on its pole energy $E_i$ to be proportional to $E_i^{1.6}$; this exponent lies between theoretical values \cite{fi86} for one- and two-body dissipation. Including all the nuclides treated in this work and the oscillator frequency ratios available from the CHFB calculations we obtain $c_w = 0.045(3)$. Of course, the proportionality constant $c_w$ has an uncertainty and its uncertainty enters in the radiative width nearly linearly as the slope of a Lorentzian sufficiently far away from $E_0$ is quasi proportional to $\Gamma_i$. In our earlier work, a value of $c_w = 0.05$ had been used \cite{ju08, ju10}; there single Lorentzians were adjusted to IVGDR data for $^{88}$Sr and $^{208}$Pb, assumed to have one pole only. The new values for $m_{eff}$ and $c_w$ given now are based on the CHFB calculations \cite{be07, de10}. With one parameter for the energies and one for the widths, both adjusted to be equal for all heavy nuclei with $A>78$ \cite{en92, ju08} one gets a good agreement to measured resonance shapes, as will be shown in Fig.~\ref{figSm} and ~\cref{figSe,figMo,figSn,figTe,figBa,figNd,figGd,figEr,figOs,figPt,figHg,figPb,figTh,figU,figIX,figTa,figAu,figPu}. We assume the width $\Gamma_i$ is varying with $E_i$ and not with $E_\gamma$, as presented previously \cite{ju08, er10, ju10}. 
 
\section{Giant dipole resonances in heavy nuclei and triaxiality}
\label{sec4}
The parameterization specified in this section has the advantage of incorporating nuclear triaxiality explicitly by setting $k=3$ in Eq. (\ref{eqdIE}). For the resulting ‘triple’ Lorentzian (TLO) description the resonance energy $E_0$ is modulated by the ratios of the oscillator frequencies ${\omega_i}$ according to Eqs. (\ref{eqPQ} and \ref{eqEi}), last line. The ratio between the two extreme resonance energies is approximately equal to $1+\beta$; the position of the middle peak is maximum for $\gamma=0$ and minimum for $\gamma=60$. This direct incorporation of triaxiality makes TLO differ from previous attempts to obtain Lorentzian fits locally using $k=1$ or 2 in Eq. (\ref{eqdIE}) for a large number of heavy nuclei \cite{ba73, be75, bo75, di88, ko90, ca09, pl11}. In many nuclei, especially those of intermediate deformation, the local fits presented there may lead to a seemingly better agreement, but often they require quite unreasonably large values for the width of the IVGDR and for the integrated strength in comparison to sum rule predictions. In section~\ref{sec6}  our earlier finding for a few nuclides \cite{ju08, er10} is now extended to many more without free fit parameters for single nuclei. We arrive at a global prescription of electric dipole strength, not using local fits for the strength and width. And this allows us to better separate the components contributing to the apparent width of the IVGDR in heavy nuclei:	
\begin{enumerate}[(a)]
  \item 	Spreading into underlying complex configurations,  	 
  \item	Nuclear shape induced splitting,  	 
  \item	Fragmentation and	
  \item	Particle escape.	
\end{enumerate}
From calculations for heavy nuclei using the Rossendorf continuum shell model \cite{ba77, be11} the escape width (d) in the IVGDR region was shown to be clearly smaller than the spreading width (a) derived by the global TLO-fit; a good agreement to data shows that $\Gamma$ in Eq. (\ref{eqEi}) is determined only by the pole energies and one global fit parameter $c_w$.  To fully understand a fragmentation (c) of the the IVGDR a quantification of various configurations is needed, as was attempted by a microscopic HFB-calculation \cite{ma16} recently, which unfortunately does not include a proper projection to discrete spin. \\

\begin{figure}[ht]
\includegraphics[width=1\columnwidth]{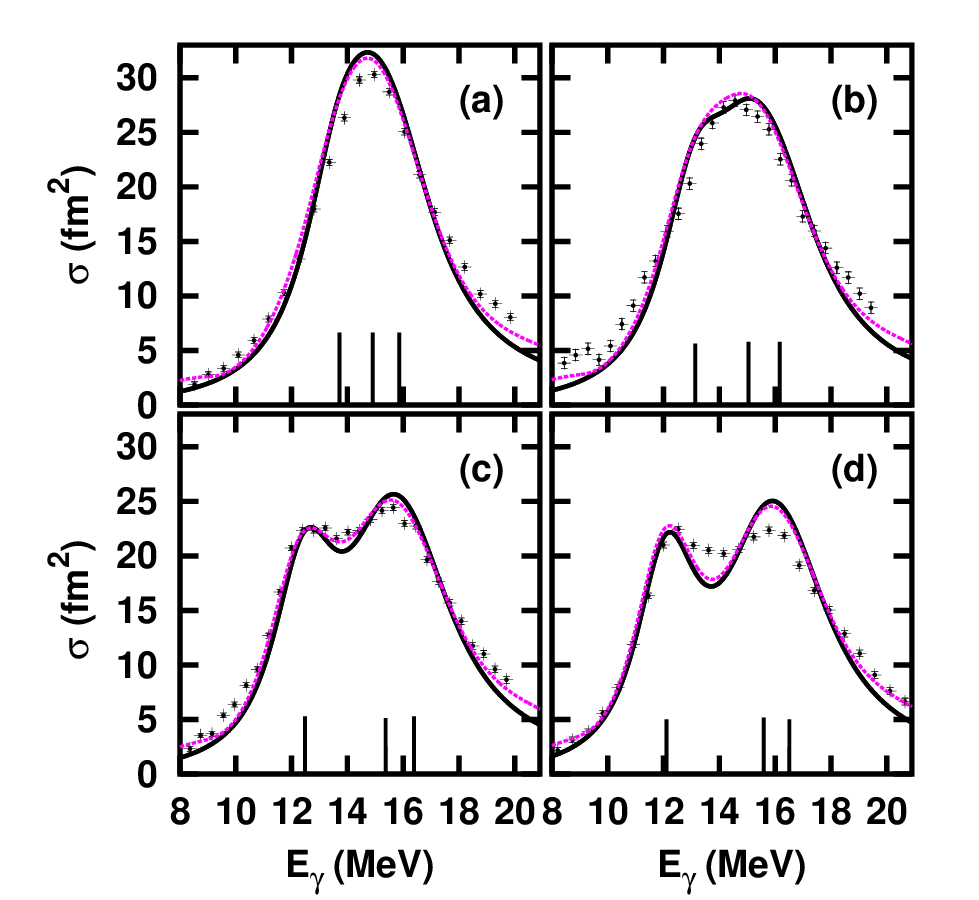}
\caption{(Color online) Photo-neutron production cross section $^{148}$Sm \cite{ca74} (a) to $^{154}$Sm (d) in comparison to the TLO sum of three Lorentzians (drawn curve) with $E_i$ indicated as black bars. The dashed (purple) curves indicate the effect of shape sampling \cite{zh09, er10, gr11}. As outlined in the next section the data were renormalized by a factor 0.9 and the calculations were folded with a Gaussian to simulate an experimental beam spread of $\sigma \cong 0.3$ MeV \cite{au70}. Panels (b) and (c) depict the situation for $^{150}$Sm and for $^{152}$Sm, respectively.} 
\label{figSm}
 \end{figure}

When a parameterization of the electric dipole strength in heavy nuclei is aimed for, the contribution of nuclear shape induced splitting (b) has to be treated sufficiently well. As proposed previously \cite{ju08, na10, er10}, a solution for this problem is found by allowing axial symmetry to be broken; this point will now be examined in further detail. As mentioned in section~\ref{sec2} accurate nuclear spectroscopic data suited to determine both deformation parameters are available only for a limited number of nuclei. The CHFB calculation \cite{de10} delivers information for the ${\omega_i}$ inserted to obtain the resonance energies in the sum of Lorentzian functions in Eq. (\ref{eqdIE}). This procedure leads to a significant splitting into three equally strong IVGDR components which increases with deformation. As will be summarized below in Fig.~\ref{figPlu}, for many nuclei the splitting between the three components is comparable in energy to their widths and thus not directly obvious from the data alone, especially in nuclides with $Q_0 \approx 200-300  {\rm fm}^2$. These are not rare as shown by the calculated triaxialities \cite{de10} (supplemental table), and we conclude that all three axes should be accounted for explicitly. This quite simple consideration explains the significant rise in apparent width as seen in Fig.~\ref{figSm} for the even Sm-isotopes with $N=86$ to $N=92$; $^{144}$Sm was not included because of the uncertain cross section for the ($\gamma$,p)-reaction.	
 The supplemental material to the CHFB-calculations \cite{de10} also gives their variances representing quantum mechanical oscillation. The respective Gaussian distributions obtained thus allow an instantaneous shape sampling (ISS) as shown earlier \cite{zh09} for isotope chains Mo \cite{er10} and Nd \cite{gr11}. There the impact of ISS on the height of the low energy tail and thus on radiative capture was demonstrated to be negligible. Results for the Sm chain as shown in Fig.~\ref{figSm} indicate a minor influence as well: the drawn black curves correspond to the TLO-prediction and the dashed purple curves stem from calculations including ISS. \\

 Special care is needed for nuclei near closed shells: The CHFB calculations do not fully account for the very deep mean field potential in such nuclei and thus they produce too much of collectivity \cite{be07}. Following what is said there, a reduction for nuclei only $\delta$ nucleons away from a shell a factor for the $\beta$-deformation \cite{de10} of $0.4 + \delta/20$ is applied for $\delta\leq 10$. This expression is used for protons as well as neutrons and the larger of the two correction factors is taken; it results in a reduction of the predicted \cite{de10} $\beta$-values by 40, 30, 20 and $10 \%$ for the isotopes $^{148}$Sm, $^{150}$Sm, $^{152}$Sm and $^{154}$Sm and the corresponding agreement to the IVGDR data is shown in Fig.~\ref{figSm}. The extension of CHFB to non-spherical nuclei introduces no extra free parameters in addition to the global ones of the Gogny-force \cite{de10}: A fit to the IVGDR energies and widths succeeds with only the four parameters introduced in Eq.(\ref{eqEi}), two of which are known from LDM mass fits. As will be demonstrated in section~\ref{sec6}  the strict distinction between damping or spreading and the deformation induced splitting allows to neglect a photon energy dependence of the width for all nuclei treated there. Thus the only local parameters for individual nuclei are the oscillator frequency ratios calculated by CHFB \cite{de10} and the widths $\Gamma_i$  in Eq. (\ref{eqEi}) vary only with the pole energies $E_i$ (and not with $A$ and $Z$). This opens the possibility for a global prediction of photon strength also for heavy exotic nuclei and has the potential of consistent predictions for radiative capture processes, where full satisfaction was not reached with presently available methods \cite{be14}. For an extension to energies below the neutron emission threshold $S_n$ modes in addition to the IVGDR have to be investigated concerning their contribution to photon absorption.

\section{Strength functions for isovector electric dipole and additional modes}
\label{sec5}
In TLO the low energy tail depends weakly on the deformation induced splitting and thus the strength there is quasi independent from any parameter. But in photon scattering experiments in this energy range narrow peaks were observed above background and these were interpreted as strength of apparently other character than isovector electric dipole \cite{ba73, kn06, he10, sa13, gr12}. In the literature the excitation energies of such modes are discussed in much more detail as compared to the strength observed. Even if it has minor importance for total photon absorption and the sum rules, it may influence the decay of the compound nucleus and consequently also the cross sections of capture reactions. Here at least their average electromagnetic strength has to be regarded. For an assessment of the agreement between TLO and experimental data the strength eventually adding to the IVGDR tail has to be characterized; if it is not of electric isovector kind it has to be described separately from IVGDR, which we propose to be in accord to TLO. As a first step of such a characterization we have derived phenomenological expressions for these 'minor' modes in their A-dependence and compare them to data, which are partly obtained at the ELBE facility at Dresden and partly derived from published work, as specified below. Following a presentation of various modes we list the obtained parameters in Table~\ref{tab1} and consider this as a basis for eventual theoretical work adressing especially strength issues. \\

A direct relation exists between ground state transition widths, summed for all levels within an energy interval $\Delta_E$, and the strength functions $f_{E\lambda}$ (in MeV$^{-(2\lambda+1)}$) defined as follows: 

\begin{align}
f_\lambda(E_\gamma) &= \frac{\sigma_{abs}^\lambda(E_\gamma)}{(\pi \hbar c)^2 g_{eff} E_r^{2\lambda-1}} \nonumber \\ 
&= \frac{1}{\Delta_E} \sum_{r} \frac{\Gamma_{r\gamma}}{E_\gamma^{2\lambda+1}} 
 \cong \frac{\langle \Gamma_{r\gamma}(E_\gamma)\rangle}{D_{r} \bar{E}_\gamma^{2\lambda + 1}}
\label{eqflam}
\end{align}

The first part of Eq. (\ref{eqflam}) relates the strength function to the photon absorption cross section ${\sigma_{abs}^\lambda(E_\gamma)}$ which is limited by sum rules. To use the average strength functions $f_\lambda (E_\gamma)$ for excitation as well as decay processes and thus connect photon scattering to radiative capture and photonuclear processes one has to suppose them to be independent of the direction of the process (Axel-Brink hypothesis)\cite{br57, ax62, ba73}. The second line of Eq. (\ref{eqflam}) directly relates $f_\lambda$ to the electromagnetic decay widths of the resonant levels $r$ in the integration interval $\Delta_E$.
Strength information can hence be obtained by summing spectroscopic width data (in MeV) over a given energy range  (also in MeV) running from $E_\gamma - \Delta E/2$  to $E_\gamma + \Delta E/2$. Their average distance is $D_{r}$ and for the sum in Eq. (\ref{eqflam}) all levels within the interval $\Delta_E$ are included; the quantum-mechanical weight factor $g_{eff}$ will be discussed later with Eq. (\ref{eqgf}).

\subsection*{Electric dipole strength below the IVGDR	}
\label{subsubsec1}
A low energy dipole mode was predicted to be formed by E3 strength coupled to low energy quadrupole modes \cite{bo57}. In many even-even nuclei rather strong photon absorption into $1^-$-levels with $1  {\rm  MeV} < E_x < 4  {\rm  MeV}$ has been observed \cite{kn95}. As a correlation of $E_x$ to the sum of the excitation energies of the low collective $2^+$ and $3^-$-modes and similar strength for odd and even nuclei was observed, this phonon coupling mode is considered well established \cite{kn95}. These photon scattering studies revealed fragmentation away from closed shells and especially in odd nuclei; it may cause a non-observation of small-strength components. To estimate the centroid energy $E_{qo}$ the sum energy of the 1st $2^+$-level \cite{ra01} and the corresponding value for the octupole ($3^-$) excitation \cite{ki80}; for exotic nuclei theoretical approximations are available \cite{de10, ro11}. \\

Already long ago ``intermediate structure'' observed by photon scattering in the energy range 5 to 8 MeV was discussed in detail \cite{ba73} and the concept of photon strength functions $f_\lambda(E_\gamma)$ was introduced to quantify its strength. For photon scattering by Zr and Sn \cite{ax70} as well as for $A \approx 200$ \cite{la79} this issue was addressed with special care in the photon detection. As mentioned there, the contribution of the quasi-continuum below the lines is significant even after the real background due to unwanted radiative processes in the detector and the near-by environment was identified and subtracted. More recently \cite{sc07, ma14} the non-nuclear scattering by the target and the near-by environment was numerically simulated, but eventually minor strength - often denoted as PDR (‘Pygmy Dipole Resonance’) \cite{ry02, sa13} - was quantified mainly by integration of the yield observed by Ge-detectors in narrow peaks.  \\

The electric dipole strength outside of the IVGDR may have isovector or isoscalar character and a distinction by experiments with isoscalar beams was proposed \cite{po92, en09, sa13}. Hints for the isoscalar character of electric dipole strength may indicate non-uniform proton-neutron distributions or compressional modes. Strength between $E_x \cong 5.5$ MeV and the neutron separation energy $S_n$ was shown to be of isoscalar nature in $^{40}$Ca, $^{58}$Ni, $^{90}$Zr and $^{208}$Pb by the coincident observation of inelastically scattered $\alpha$-particles and de-excitation $\gamma$-rays \cite{po92}. In the present study the various results reviewed recently \cite{sa13} for this energy range are tentatively separated into two components and named low and high energy pygmy mode (PM).  The parameters given in Table~\ref{tab1} for the strength integrals should be understood as approximate. As becomes obvious in subsequent figures, the low energy PM (a seemingly resonant strength near $0.4  \times  E_{IVGDR}$) appears to be of similar magnitude for many different $A$, when compared to the IVGDR. Experimental ``evidence for a 5.5-MeV radiation bump'' in nuclei near Pb, an intermediate structure named ‘pigmy’ \cite{ba73} 40 years ago, was recently extended \cite{ma14} to isotopes in the N = 50 and N = 82 region, and this suggests our nomenclature as high energy PM.  

\subsection*{Magnetic Dipole Strength}
\label{subsubsec2}
Magnetic (M1) strength is weaker as compared to E1 \cite{ko90} and , as outlined in a recent review \cite{he10}, its spin flip component occurs at higher energy than collective orbital magnetic strength (scissors mode). The latter is strong in nuclei with a large quadrupole moment and high resolution photon scattering data \cite{sc88, sc07, la87, en09, ru13} show the strength of  M1 transitions to be below that for E1 for energies above 2 MeV. The magnetic strength in Table~\ref{tab1} is mainly derived from a review published recently \cite{he10}. 
\begin{table*}
\begin{tabular}{llllr}
 \hline
 Component  & multi- \\ parameter(units) & polarity &  $E_c$ (MeV) & $I_c$ (fm$^2$MeV) & $\sigma_c$ (MeV)\\
 \hline
 low $E_x$ pygmy mode &  E1 & $0.43 E_0$ & 7 $ \frac{Z (N-Z)}{A}$ & 0.6\\
high $E_x$ pygmy mode &  E1 & $0.55 E_0$ & 13 $ \frac{Z (N-Z)}{A}$ & 0.5\\
$0^+ \leftrightarrow (2^+\times3^-)_{1-}$ & E1 & $\frac{140}{N}(1+\frac{107}{Z})$ & $0.006ZA\beta$ & 0.6\\
orbital  (scissors) mode & M1 & $0.21 E_0$ & $0.033ZA\beta$  & 0.4\\
isoscalar spin-flip & M1 & $42 A^{-1/3}$ & 17 & 0.8 \\
isovector spin-flip & M1 & $47 A^{-1/3}$ & 27 & 1.3 \\
low $E_x$ quadrupole & E2 & $19 A^{-1/3}$ & 0.1 $\frac{\alpha (\pi R_p E_\gamma)^2 Z^2}{(3 A m_p c^2)}$ & 1.0 \\
ISGQR & E2 & $63 A^{-1/3}$ & 2 $\frac{\alpha (\pi R_p E_\gamma)^2 Z^2}{(3 A m_p c^2)}$ & 1.0 \\
IVGQR & E2 & $48 A^{-1/6}$ & $\frac{\alpha (\pi R_p E_\gamma)^2 Z^2}{(3 A m_p c^2)}$ & 1.8 \\
\hline
\end{tabular}
\caption{ Parameters of an upper limit for three minor electric and three magnetic modes contributing to the dipole strength function to be calculated with a Gaussian (Eq.\ref{eqfGaus}). For a Gaussian with a standard deviation $\sigma_c$ a shape near the peak is reached very similar to a Lorentzian with the width $\Gamma$, which is larger by a factor 2.5 as compared to $\sigma_c$. Three GQR modes are also listed, which are seen in photon absorption as well; they are proportional to the square of the nucleus’ charge radius $R_p$ (rms).}
\label{tab1}
\end{table*}

\subsection*{Electric quadrupole modes}
\label{subsubsec3}	 
Low energy E2-transitions can be considerably enhanced and have played an important role in the spectroscopy of heavy nuclei – as discussed in \cref{sec2}. The photon absorption cross sections can be derived from Eq. (\ref{eqGr0}); they have been found to be small as compared to E1 absorption \cite{we80} in the energy region near $S_n$. This is also true for the absorption into the third $2^+$-state predicted for triaxial nuclei \cite{df58, de10} to have a B(E2) value up to $ 7 \% $ as compared to the first. Extending previous work \cite{we80, ju08, ca09} the influence of quadrupole giant resonance (GQR) contributions to photon absorption was investigated by us; information for quadrupole strength comes from sum rules and theoretical predictions \cite{na65, we80, wo91} adjusted to electron scattering data \cite{pi74, sc75, do82}. Strength modifications at higher energy due to the isovector IVGQR appear to be important, as seen in the upcoming \cref{figSe,figMo,figSn,figTe,figBa,figNd,figGd,figEr,figOs,figPt,figHg,figPb,figTh,figU,figIX,figTa,figAu,figPu}. Here $f_{E2}$ was multiplied with $E_\gamma^2$, the additional energy dependent phase space factor for absorption and decay, to allow a visual comparison to $f_1$; for the calculation of the integral $I_{E2} = I_c $ (as listed in Table~\ref{tab1}) the correct decay width is used in Eq. (\ref{eqIr0}). The isoscalar ISGQR lies not far from the pole of the IVGDR such that it adds to the strength observed there, causing a small deviation from TLO. \\

  The approximations used in Table~\ref{tab1} for the integrals of the E1, E2 and M1 components were derived through a comparison to respective observations \cite{la87, ko90, he10} under respect of the above mentioned information. The resulting parameters for these ``minor'' contributions to the photon strength function are listed there; independent of an interpretation of these modes a Gaussian seems justified to describe them. No arguments were found, why Lorentzians should be preferred; this differs from the IVGDR, where a Lorentzian with energy independent width is appropriate \cite{do72, go77}. Separate Gaussian distributions	 	
 \begin{equation}
f_\lambda(E_\gamma)=\frac{1}{(\pi\hbar c)^2 g_{eff} E_\gamma^{2\lambda-1}} 
\frac{I_c}{\sqrt{2\pi\sigma_c^2}}\exp{\left(\frac{-(E_\gamma-E_c)^2}{2\sigma_c^2}\right)}
\label{eqfGaus}
\end{equation}
for each of them are used and this supresses unobservable strength in long tails; the parameters were selected to somewhat overpredict available data after the TLO-integral over the energy interval is subtracted \cite{ma14}. Table~\ref{tab1} lists the cross section peak integrals $I_c$ as well as the centroid energies $E_c$ and the standard deviations $\sigma_c$ (in MeV) as discussed above. The $I_c$ in Eq. (\ref{eqfGaus}) as well as in Table~\ref{tab1} are sums over many individual levels within the Gaussian like in Eq. (\ref{eqdIE}) for the IVGDR and at variance to Eq. (\ref{eqsig}), valid for a single level. In the next sections the results corresponding to a sum of TLO and the 6 Gaussians for the "minor" modes are shown as blue lines in \cref{figSe,figMo,figSn,figTe,figBa,figNd,figGd,figEr,figOs,figPt,figHg,figPb,figTh,figU,figIX,figTa,figAu,figPu}. The average quantity $f_\lambda$ is depicted, which usually \cite{ba73} is assumed to be insensitive to details of the nuclear excitations and hence one approximates collective electromagnetic transition strengths of energy $E_\gamma$ = $E_r$ - $E_f$  to be independent of the energies $E_r$ and $E_f$  by using $f_\lambda(E_\gamma)$; together with the notion of $f_\lambda$ being valid for excitation and decay (Axel-Brink hypothesis) . It should be stressed, that the contribution of all "minor" modes to the sum rule integral in Eq. (\ref{eqIE1}) is weaker by at least one order of magnitude as compared to the IVGDR sum. The relative importance of the components for cross sections calculated as presented here may be found in a letter published recently \cite{gr14}. There approximate expressions for the level density were combined to TLO and the minor modes to derive a global comparison to experimental radiative neutron capture cross section data; the assumption of broken axial symmetry considerably influenced the calculated level densities and improved our predictions.

\section{Electromagnetic strength in the IVGDR and below}
\label{sec6}
The TLO ansatz and the presented phenomenological description of 'minor' strength  do not aim for a full theoretical understanding of the coupling between the IVGDR to quadrupole modes or other excitations, but only for a prediction of photon strengths, which is global and hence extendable to many nuclides. In view of the already previously observed \cite{ba73} ''difficulty of accounting for the bump with the aid of a smoothly varying strength function", we clearly distinguish between the Lorentzian tail and ''minor" intermediate strength. Such an ansatz was successfully applied to near (semi-)magic nuclei \cite{ax70,la79}, but later opposed heavily \cite{ko90, ko93, mu00}. Here fits on the basis of the KMF model \cite{ka83} were applied to data for nuclei with larger deformation, but we show that using TLO, {\em  i.e.} giving up axiality, describes existing data at least equally well. Most of them were obtained with quasi-monochromatic photons from positron annihilation in flight \cite{be75, di88}, but such data do not exist for all stable isotopes and for some nuclei the total photon cross section has been studied only by an absorption technique. These data may serve for a consistency check in spite of systematic errors due to the need to subtract the strong atomic absorption. \\

Experimental photo-neutron or photon absorption cross sections were published for various nuclei and are available electronically \cite{ex14}. The upcoming discussion presents in
\cref{figSe,figMo,figSn,figTe,figBa,figNd,figGd,figEr,figOs,figPt,figHg,figPb,figTh,figU,figIX,figTa,figAu,figPu} a selection data for the IVGDR peak region. They have been averaged by rebinning and eventually corrected for several facts: 
\begin{enumerate}	
\item Photo-neutron data were often obtained by using quasi-monochromatic photon beams with a rather wide energy distribution, which is incorporated by folding the calculations with a Gaussian of width  $\sigma$= 0.3 MeV, a value not as large as some recently made guesses \cite{va04, va14}. Also in the case of a bremsstrahlung distribution used as photon source quite some uncertainty may arise \cite{fi83}; in some cases energy calibrations may differ somewhat. 
\item Considerable discrepancies were reported for experiments performed {\em  e.g.} at different laboratories and
it has been found by various studies \cite{ca71, be71, be87, is04, va14, sa14, ny15}, that photoneutron cross sections determined at Saclay around 1970 should be reduced. The necessary reduction is probably related to difficulties in the analysis of multi-hit events in the neutron detector array \cite{va14, is04}; this also influences the correction for the ($\gamma$,2n) channel at higher energy.  In accordance to a precision study \cite{be87}, confirmed by results \cite{na08,er10} from the radiation source ELBE, the photo-neutron data of that origin are hence multiplied in this work by 0.9, considered as suitable for various $A$.  
\item  Most of the targets used contain isotopic contaminations, and when some of them have a lower $S_n$ (like many odd isotones) the low energy yield has to be discarded or corrected \cite{er10}.	
\item Below and above the pole of the IVGDR contributions from the giant quadrupole (GQR) modes may appear; sufficiently accurate GQR-data could not be found and approximate assumptions have been discussed above together with observed photon strength below the neutron threshold which adds to the IVGDR tail. Here Porter-Thomas fluctuations \cite{ax62} may randomly create strong peaks in spectra from a quasi-continuum of weakly populated levels. 
\end{enumerate}

Photon strength data for energies below $S_n$ have been published but not always a completeness on absolute scale was assured \cite{ju08, he10, sa13}. Results from photon scattering have to be corrected for branching to other than the ground state and in various cases the bremsstrahlung continua also feed higher excited levels and their decay yield has to be subtracted. For a globally applicable quantification the yield not seen as individual spectral lines in a Ge-detector (looking like quasi-background) has to be taken into account \cite{sc07, ma14}. In the case of less complex spectra use can be made of data from the decay to well isolated levels \cite{ju95, go98}. Data analysis schemes were developed {\em  e.g.} at ELBE \cite{ru08, er10, sc12} which are based on statistical considerations already formulated some time ago \cite{ba73, ax70}. Due to the fact, that the electromagnetic strength is responsible for the absorption as well as the emission of photons, an iterative procedure leads to a self-consistent solution \cite{sc12, ma12, ma14}. Here, nuclear level densities enter which we now take from a new formulation \cite{gr14} which includes a collective enhancement due to the breaking of axial symmetry. In the case of some ELBE-data this lead to a reduction of $30\%$ in the $f_\lambda$ as compared to previously published work and this is mentioned in the respective figure captions. In principle, also photon yields observed after nuclear reactions can deliver strength-information, when they are normalized via an "external" fixed point. In the case of resonant neutron capture this is realized by "using the absolute gamma-ray intensities due to captured thermal and resonance neutrons" \cite{mu00}. But the observed spectra \cite{be95} indicate, that the strength published was observed only because it was significantly stronger than the one buried in the experimental background ({\em  e.g.} due to Compton scattering in the photon detector and its environment). 

In
\cref{figSe,figMo,figSn,figTe,figBa,figNd,figGd,figEr,figOs,figPt,figHg,figPb,figTh,figU,figIX,figTa,figAu,figPu}
photon strength function data are compared to the TLO parameterization with three IVGDR pole energies induced by the non-axiality \cite{de10} and the global parameters in Eq. (\ref{eqEi}). The (black) dashed lines represent the prediction of $f_{E1}$ thus derived with the resonance integral from Eq. (\ref{eqdIE}) (in accordance to the TRK sum rule) equally divided among the three poles of TLO. In all these figures the three are indicated as black bars at the energy axis. It is worth mentioning that  the absolute heights in the low energy slope are nearly unchanged by the splits, albeit the apparent peak-height depends on it. As shown previously \cite{sc12, gr14}, the effect of the photon strength on radiative neutron capture is strongest in the tail region approximately at $50\%$ of the neutron binding energy $S_n$ as the folding of the level density in the final nucleus with the photon strength function becomes especially large. Up to eight additional "minor" strength components may be of importance there; they have been detailed in section~\ref{sec5}. For a specification of their energy, strength and width we have used published data as listed with the figures. But the available information is by far not detailed enough to derive a systematic parameterization to evaluate their eventual influence on neutron capture with high precision. This is demonstrated by the experimental data inserted in the figures; the result of our adjustment of the relevant parameters to derive an approximate agreement is depicted by a full (blue) line. Like for TLO only globally fitted quantities enter the calculations for these plots; they are compiled in Table~\ref{tab1}. The plots start at 3 MeV as below Thomson scattering by the nuclear charge surmounts the IVGDR tail and especially the zero at $E_\gamma = 0$ in Eq. (\ref{eqdIE}). \\

For the nuclei to be discussed photo-neutron data and results for energies below $S_n$ are available in a wide range of $A$, $Z$ and $Q_0$. A comprehensive collection of photon absorption data are available from the EXFOR data base \cite{ex14}, for which they were extracted from original work. To improve the visibility of the data points, some of the excitation functions have been re-binned to around 0.6 MeV/bin, and points of no significance were suppressed, {\em  e.g.} when their uncertainty is comparable to the value. The subsequent figures will demonstrate what features of photon strength can be derived experimentally, and how well systematic trends become visible. For nuclides with $A< \approx 60$ channels competing to ($\gamma$,n) may influence the extraction of the photo-nuclear cross section \cite{di88} and hence the photon strength; such nuclei are not discussed in this study. 

\subsection*{Even-even nuclei}
\label{subsubsec4}
\begin{figure}[ht]
\includegraphics[width=1\columnwidth]{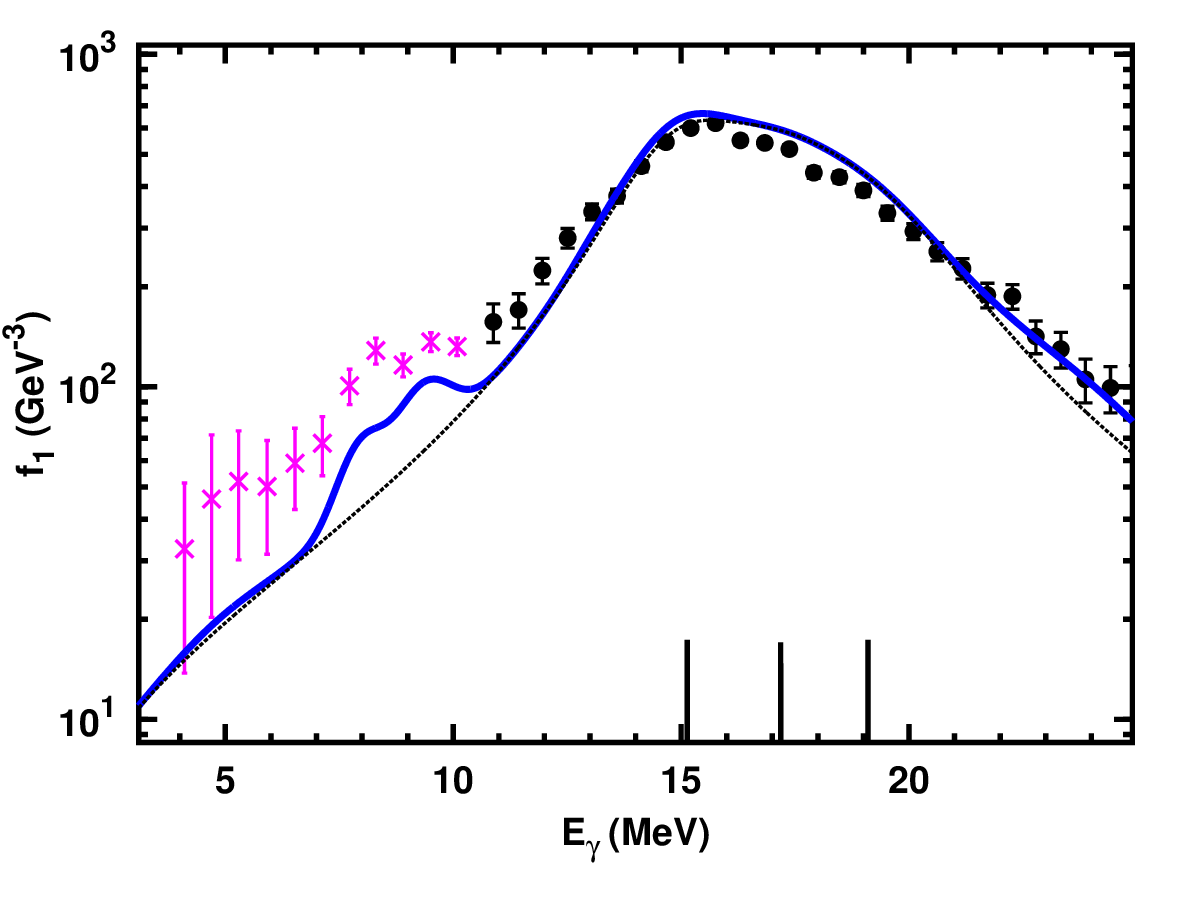}
\caption{(Color online) Photon strength function for $^{78}$Se: the TLO-prediction is depicted as dashed black line and a full line in blue represents the approximate influence of minor strength. Photoneutron results  (black dots \cite{ca76}) are shown in comparison as well as photon scattering data at low $E_\gamma$; these (x-symbols in magenta) were obtained with bremsstrahlung from ELBE \cite{sc12,ma14} and reduced by $30 \%$ in view of a new level density ansatz  \cite{gr14}. } 
\label{figSe}
 \end{figure}

In $^{78}$Se a significant increase over the extrapolated tail is observed from photon scattering investigated at the radiation source ELBE for $7<E_\gamma<10 $ MeV, although these data \cite{sc12} were reduced by $30 \%$  in Fig.~\ref{figSe}. This is a correction for the rescaled level density ansatz used in the data analysis  \cite{gr14}  as compared to the one published recently \cite{sc12}. In any case, there is an overshoot above TLO, which is partly accounted for by our phenomenological ansatz for 'minor' strength.	

\begin{figure}[ht]
\includegraphics[width=1\columnwidth]{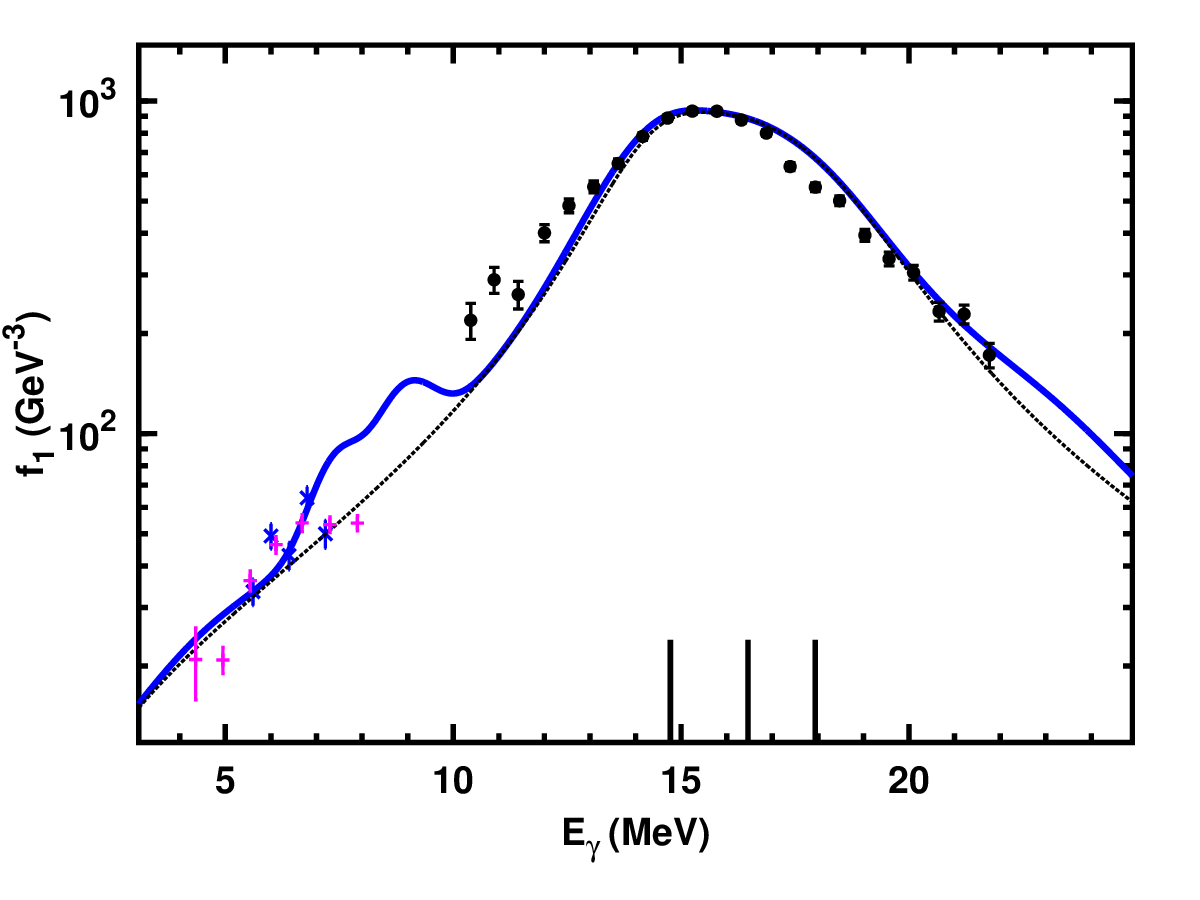}
\caption{(Color online) Photon strength for $^{98}$Mo from photoneutron data \cite{be74} (black circles) in comparison to the sum of three Lorentzians (TLO) as described for Fig.~\ref{figSe}. The data below 9 MeV are from elastic photon scattering $^{98}$Mo($\gamma,\gamma$) observed with monochromatic photons \cite{ru09} (blue x) or bremsstrahlung \cite{ru08} (magenta $+$ symbols, modified by 0.7). } 
\label{figMo}
 \end{figure}

For $^{88}$Sr \cite{ju08,gr14}, as well as for $^{92-100}$Mo \cite{ru08, er10} minor strength was observed at low energy similar to $^{78}$Se. There the experiments and their analysis as performed at the ELBE facility are described in detail. The similarity between the low energy slopes in the experimental data of all Mo isotopes led to the suggestion \cite{ju08, er10} of an extrapolation with energy independent width under the assumption of broken axial symmetry. In Fig.~\ref{figMo} for $^{98}$Mo bremsstrahlung data corrected statistically for inelastic scattering \cite{ru08, er10} in the tail below 10 MeV are very close to newer results obtained with quasi-monochromatic photons \cite{ru09}.	
At variance to this agreement photon strength function data derived from gamma decay after inelastic $^{3}$He scattering \cite{gu05} are nearly a factor of 3 smaller at 7 MeV as compared to the data as shown here and this may indicate, that the so-called Oslo method may suffer from using uncertain level density information, which may have to be modified as mentioned above. The recently revised data \cite{la10} agree around 7 MeV with our photon scattering results and thus also with the TLO prediction with minor strength added. 

\begin{figure}[htb]
\includegraphics[width=1\columnwidth]{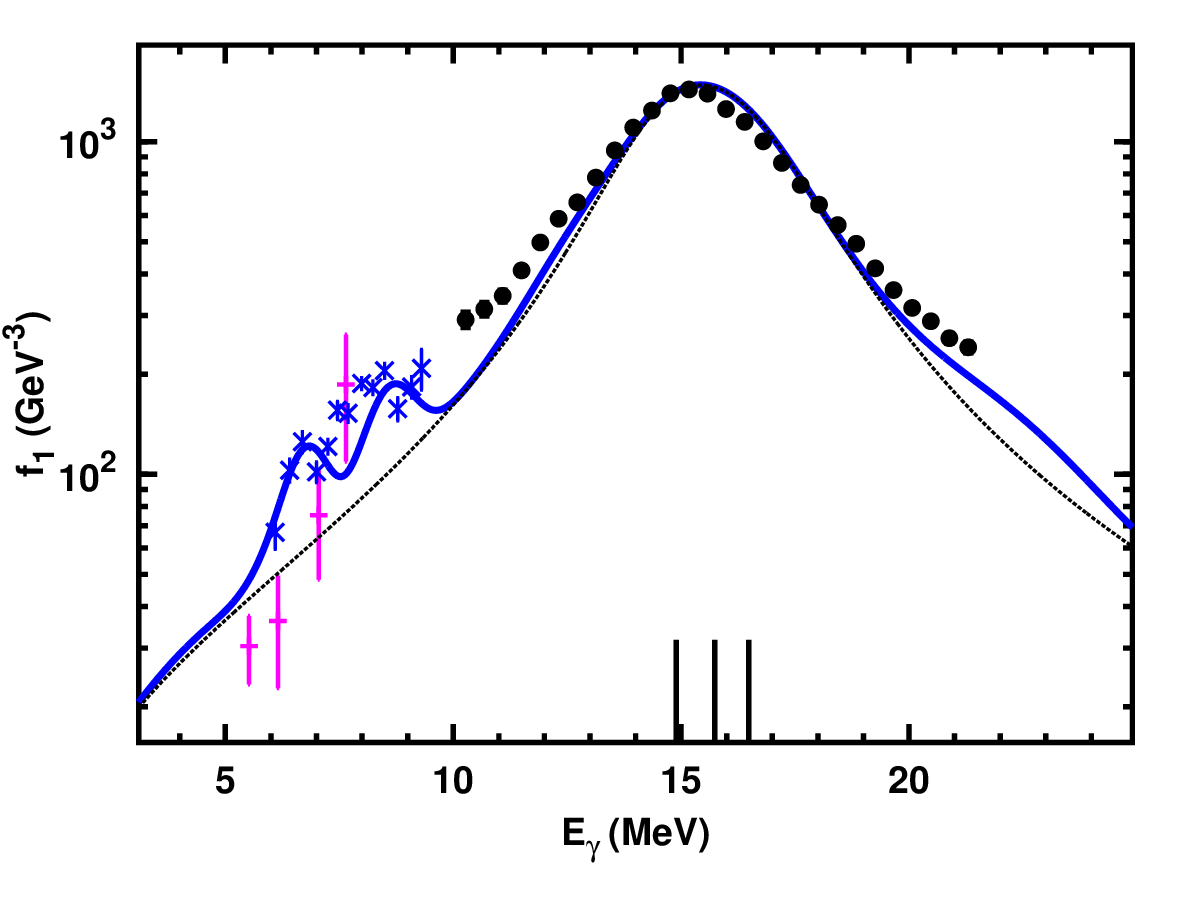}
\caption{(Color online) Photon strength for $^{118}$Sn from photon scattering \cite{ax70} (blue x symbols ) at low and photo-dissociation \cite{le74} (black circles) at higher energy in the IVGDR range. The data are compared to TLO, with deformation reduced to $\beta=0.066$, which is below the value derived from B(E2,${0^+ \rightarrow 2^+}$) \cite{ra01} (see caption of Fig.~\ref{figSe} and text near it). Recently reanalyzed data \cite{to11} from gamma decay after $^3$He induced reactions are shown as well ($+$ symbols in magenta with large uncertainty bars).} 
\label{figSn}
 \end{figure}

Quasi-elastic photon scattering from natural Sn has been studied long ago \cite{ax70} at the tagging set up installed at Urbana and "intermediate structure" in addition to the IVGDR tail has been identified. The absorption cross sections were derived from scattering data \cite{ax70}; their branching correction by inserting constant average resonance widths may overestimate $\sigma_{abs}$ by at most $20 \%$ \cite{ax70}. To improve the overlap with the ($\gamma$,n)-data a reduction by 0.8 was applied in Fig.~\ref{figSn} and the resulting values are shown together with the cross section for $^{118}$Sn ($\gamma$,xn) \cite{le74}, obtained with positron annihilation in flight; the surprisingly large strength near 10 MeV may be related to a target admixture of odd isotopes, similar to what was worked out for Mo targets \cite{er10}. Recent experiments with laser backscattered photons \cite{ut11}, which cover the threshold region, support such an assumption for $^{118}$Sn. 	

From a high resolution photon scattering experiment \cite{go98} with correction for branching losses a strength enhancement near 6.5 and 8 MeV was reported for $^{116}$Sn and $^{124}$Sn, similar to what is shown here for $^{118}$Sn. A recent study of $^{112}$Sn and $^{120}$Sn \cite{oe14} uses statistical corrections for inelastic scattering as proposed earlier \cite{ax70, ru08}; from a fluctuation analysis they propose to increase the final photon strength by nearly a factor of two with respect to the sum of peaks observed by Ge-detectors with high resolution. Recently reanalysed data on $f_\gamma$ obtained at Oslo for $^{118}$Sn \cite{to11} agree to older ones \cite{ax70} - as depicted in Fig.~\ref{figSn}. The overshoot seen at the higher energies can be explained by the influence of the giant quadrupole resonance and is of no importance for the discussion here. \\

\begin{figure}[ht]
\includegraphics[width=1\columnwidth]{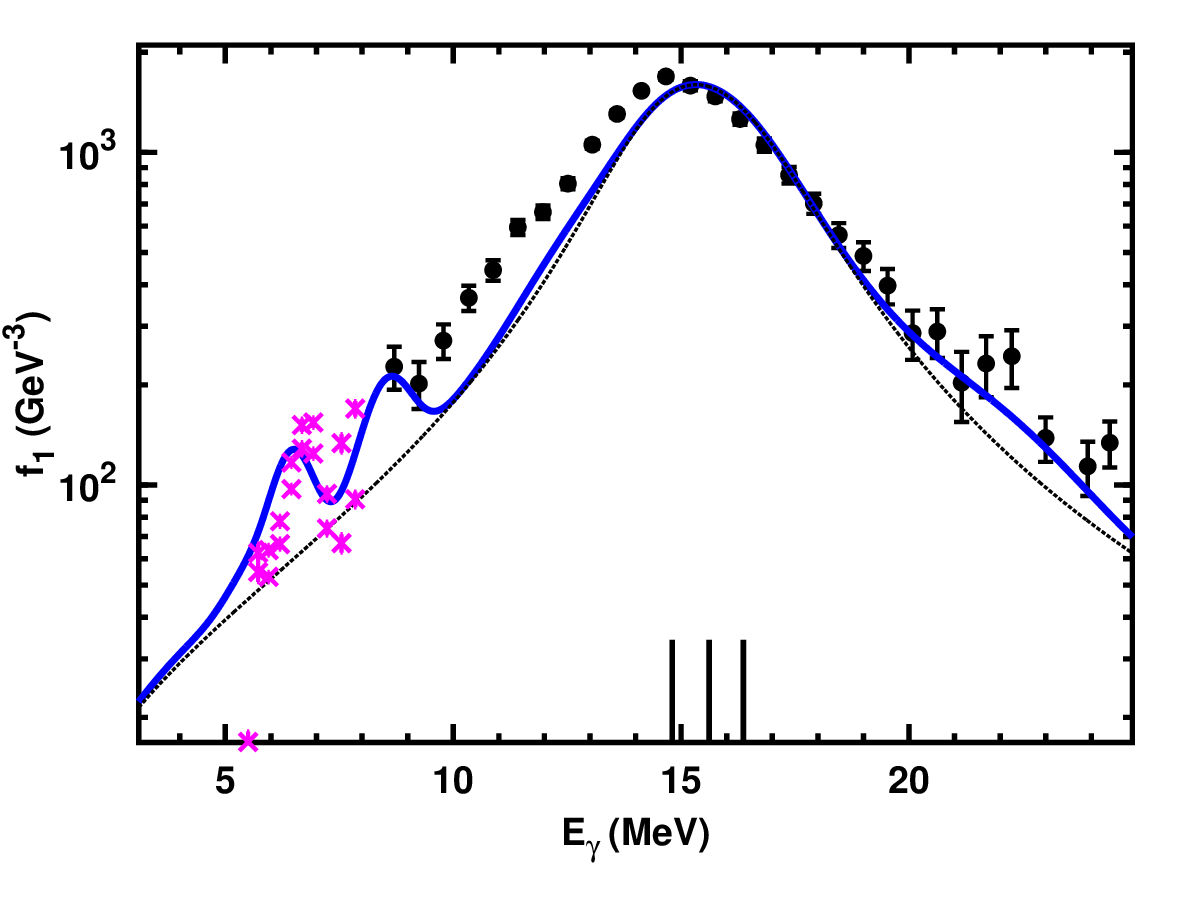}
\caption{(Color online) Photon strength for $^{130}$Te calculated as sum of three IVGDR-Lorentzians (TLO)(see the caption of Fig.~\ref{figSe} ) in comparison to data from photoneutron production in natural Te \cite{le74} (black circles). The data below $S_n$ (x, magenta) are from a careful analysis \cite{is13} of scattering data taken with quasi-monoenergetic photons, yielding lower and upper limits for the photo-absorption cross section. } 
\label{figTe}
 \end{figure}

\begin{figure}[ht]
\includegraphics[width=1\columnwidth]{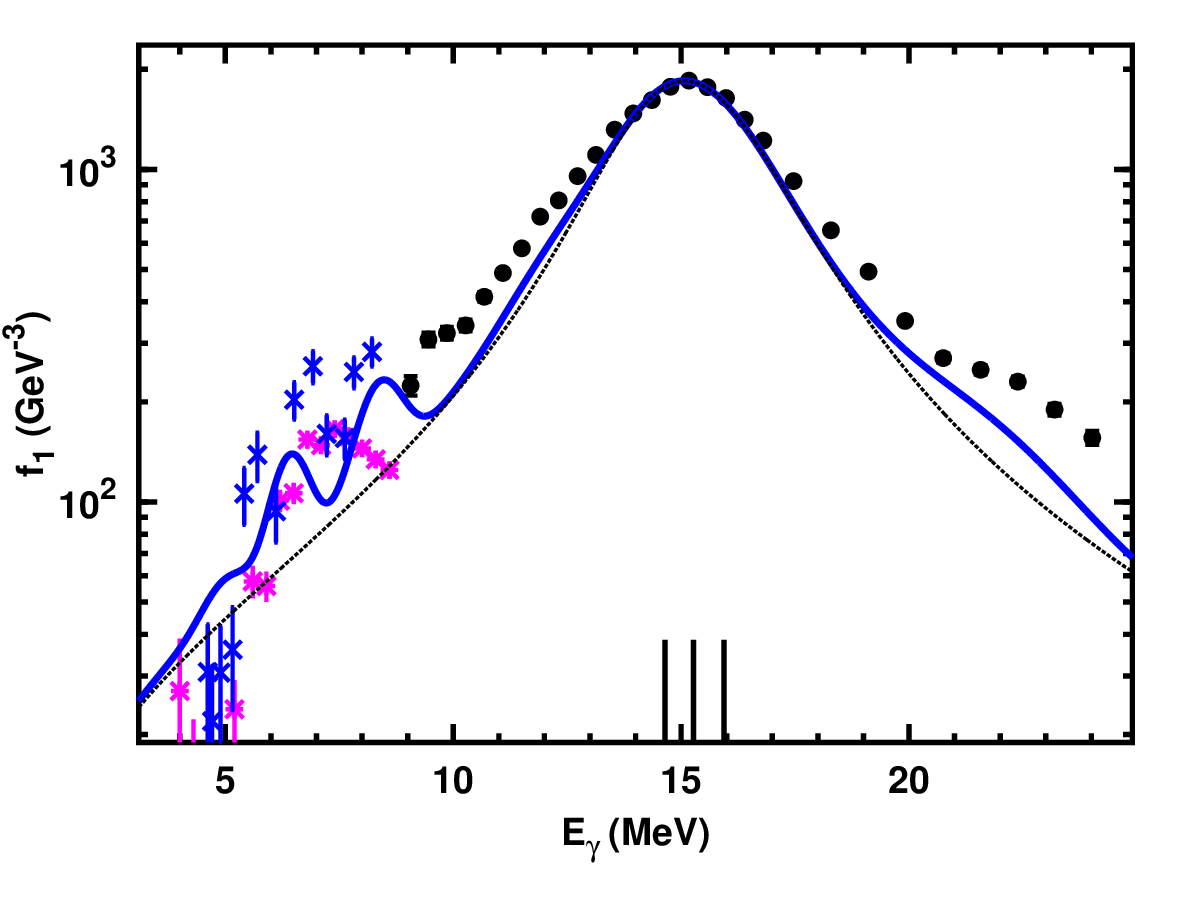}
\caption{(Color online) Photon strength  derived by photo-neutron production on $^{nat}$Ba \cite{be71}(black circles) in comparison to TLO for the IVGDR for $^{138}$Ba (see the caption of Fig.~\ref{figSe} ). Photon scattering data for $E_\gamma < S_n$  are shown as blue x-symbols \cite{to10} and magenta asteriks for $^{136}$Ba \cite{ma12} reduced by $30 \%$. } 
\label{figBa}
 \end{figure}

In Fig.~\ref{figTe} results from photo-neutron emission from $^{nat}$Te, multiplied by 0.9 – as done in general for data from Saclay – agree well to TLO above the IVGDR peak, which may be widened by the various isotopes in the target. On the low energy side of the peak the isoscalar component of the GQR is expected from the systematics for this mode; its influence on the photoneutron cross section is not completely clear. The low energy data support the finding of "intermediate structure" or "pygmy" strength as observed since long \cite{ax70, ba73, la79} near $0.4 E_0$, also seen in Figs.~\ref{figSn} and \ref{figBa} and to some extent in the data below 10 MeV for all nuclei presented here. 

For $^{138}$Ba scattering data have been taken with quasi mono-energetic photons at the laser backscattering beam at HI$\gamma$S \cite{to10}, whereas a bremsstrahlung experiment at ELBE was performed with $^{136}$Ba \cite{ma12}. As one can assume that the two data sets result in similar absorption cross sections, they are shown together in Fig.~\ref{figBa}. 

For the nuclides to be discussed in the following ($^{146}$Nd to $^{190}$Os) photon strength information for  $E_\gamma < S_n$  was obtained from individually known branching ratios of gamma transitions following neutron capture via resonances near $S_n$   \cite{ba73} or by analyzing gamma spectra following average resonance capture (ARC) \cite{mu00} to reach the nucleus in question. Experimentally  an inspection of the gamma-ray angular distributions assures $\lambda=1$, and the decay multipolarity (E1 or M1) is derived by using several neutron energies \cite{ko90, ko93}. Inserting these widths and the average level spacings into Eq. (\ref{eqflam}, second line) results in $f_\lambda(E_\gamma)$, but this relies on a known level density $1/D_r$ \cite{ko90, ko93, be95, ko08, eg09, gr14}. 

\begin{figure}[ht]
\includegraphics[width=1\columnwidth]{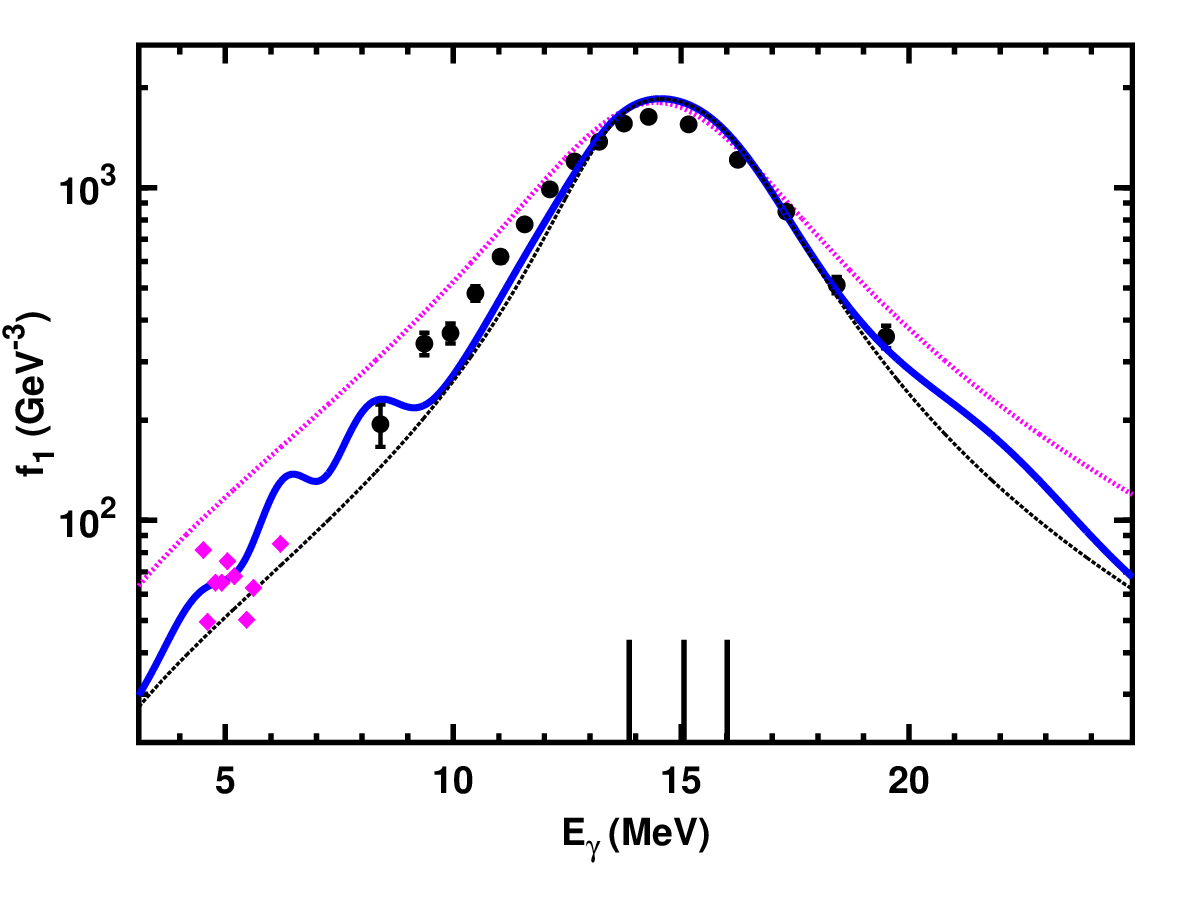}
\caption{(Color online) The photon strength for $^{146}$Nd in comparison to TLO for the IVGDR (dashed line in black). The data above $S_n$ are from photo-neutron production \cite{ca71} and the ones below (magenta diamonds) are derived from gamma decay subsequent to ARC by $^{145}$Nd \cite{mu00}. The dotted line in magenta results from a fit to these data with one pole only \cite{pl11} and the comparison including minor strength is depicted as blue line (see caption of Fig.~\ref{figSe})} 
\label{figNd}
 \end{figure}
 
In Fig.~\ref{figNd} the case of  $^{146}$Nd is shown and the triaxiality in TLO leads to a reasonable description of data in the region of the IVGDR as well as below. In accord to Eq. (\ref{eqdIE}) and (\ref{eqEi}) comparatively small $\Gamma_i$ of 2.82, 3.33 and 3.76 MeV are used without a decrease with $E_\gamma$. In  previous work \cite{be75, ca09, pl11} a single Lorentzian (SLO, $k=1$) was proposed for $^{146}$Nd together with  $\Gamma_{IVGDR}=5.74$ MeV, also shown in Fig.~\ref{figNd} as magenta curve. It indicates that such a fit leads to a large resonance width $\Gamma$, as it emphasizes the pole region. Only with a decrease of the IVGDR width with photon energy, as was assumed \cite{ko90, be95, kr04, ca09} in KMF-type analyses \cite{ca09}, agreement to the low energy data near 5 MeV is reached in spite of the large value for $\Gamma$.  For even lower energy it was proposed to add a component to the Lorentzian which violates the Axel-Brink hypothesis; this has allowed to assign E1-character to the data from $^{143}$Nd(n, $\gamma \alpha$)  \cite{po72, al77, po82} with their strength observed even below 1 MeV. But this neglects the fact, that M1 radiation should be favored \cite{fu73}, similar to what was indicated theoretically for $^{94-96}$Mo \cite{sc13}. \\
 
\begin{figure}[ht]
\includegraphics[width=1\columnwidth]{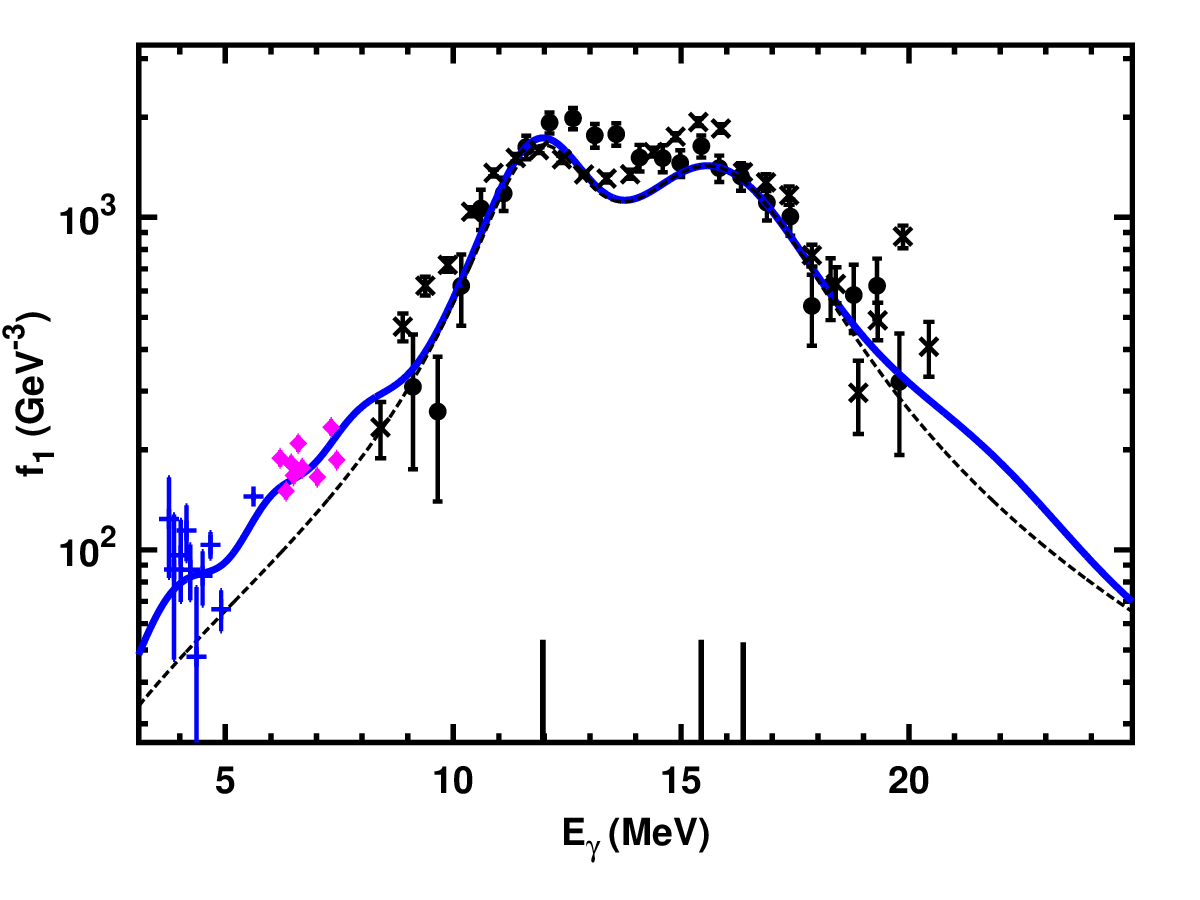}
\caption{(Color online) Photon strength for $^{156}$Gd in comparison to TLO (see the caption of Fig.~\ref{figSe} ). Data for 6-8 MeV (\cite{mu00} magenta diamonds) are derived from ARC as well as those below, which are for $^{157}$Gd (\cite{ko93} blue +). The photon absorption (\cite{gu81} black circles) and photo-nuclear data (\cite{va69} black x-symbols) do not fully agree in the peak region.} 
\label{figGd}
 \end{figure}

\begin{figure}[ht]
\includegraphics[width=1\columnwidth]{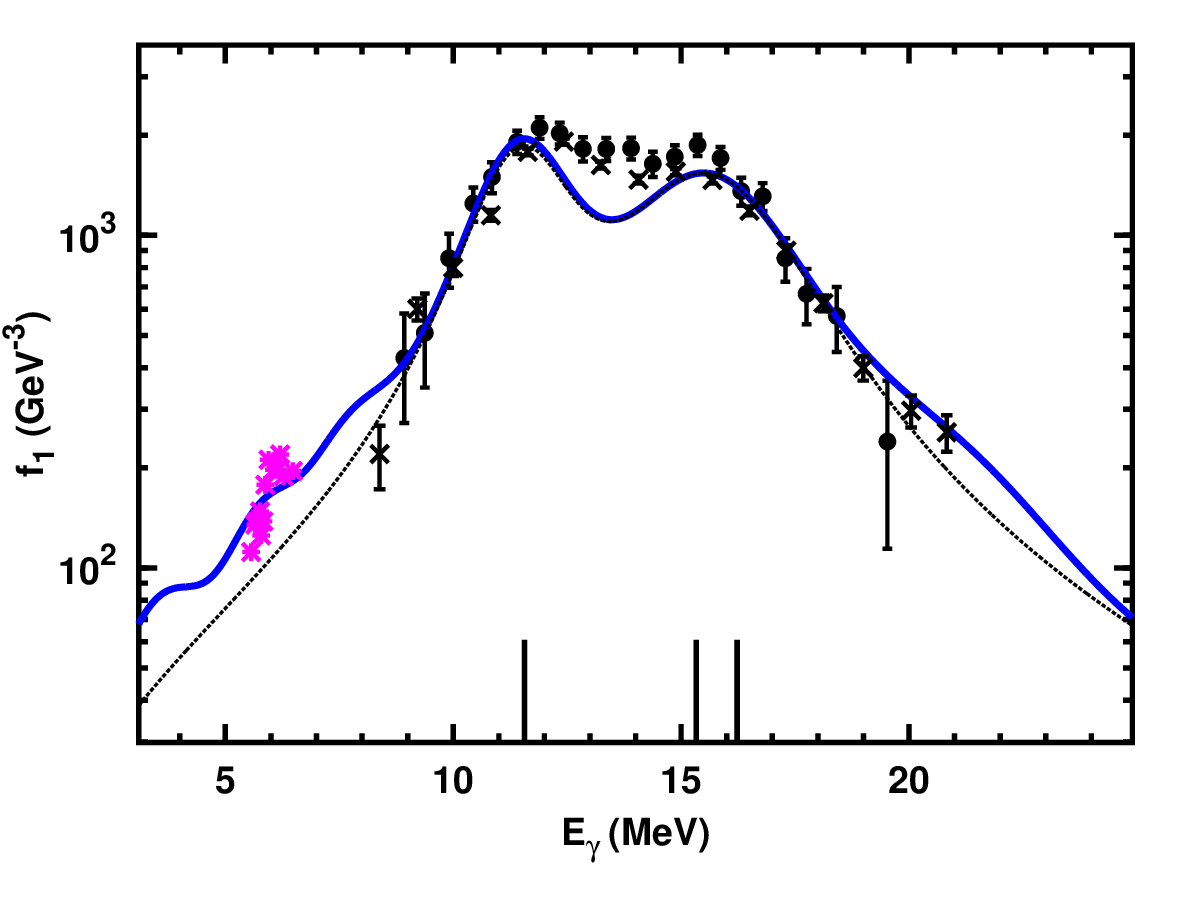}
\caption{(Color online) Photon strength for $^{168}$Er derived from ARC data (magenta * symbols \cite{mu00}) and from the cross section of photon absorption (black dots \cite{gu81}) in comparison to TLO (dashed curve in black). Shown as full line in blue is the prediction with minor strength and photo-neutron data for $^{\rm nat}$Er($\gamma$,xn) \cite{be68} are depicted as black x-symbols.} 
\label{figEr}
 \end{figure}

Figs.~\ref{figGd} and \ref{figEr} show results of experiments for the IVGDR range in $^{156}$Gd and $^{168}$Er; they were selected as an example for experimental uncertainties to be aware of in discussions about details of the dipole strength: The data with the larger error bars were obtained by photon absorption with subsequent subtraction of the strongly dominating absorption by the atomic shell, which has to be determined by a precise calculation. The other data are from  photo-neutron experiments; the deviation in the minimum near 14 MeV may result from a large energy width of the photon beam, but still the agreement to TLO is remarkable. In view of the large Porter-Thomas fluctuations, which have been averaged out only partly, this also holds for the low energy data after 'minor' strength is added.

\begin{figure}[ht]
\includegraphics[width=1\columnwidth]{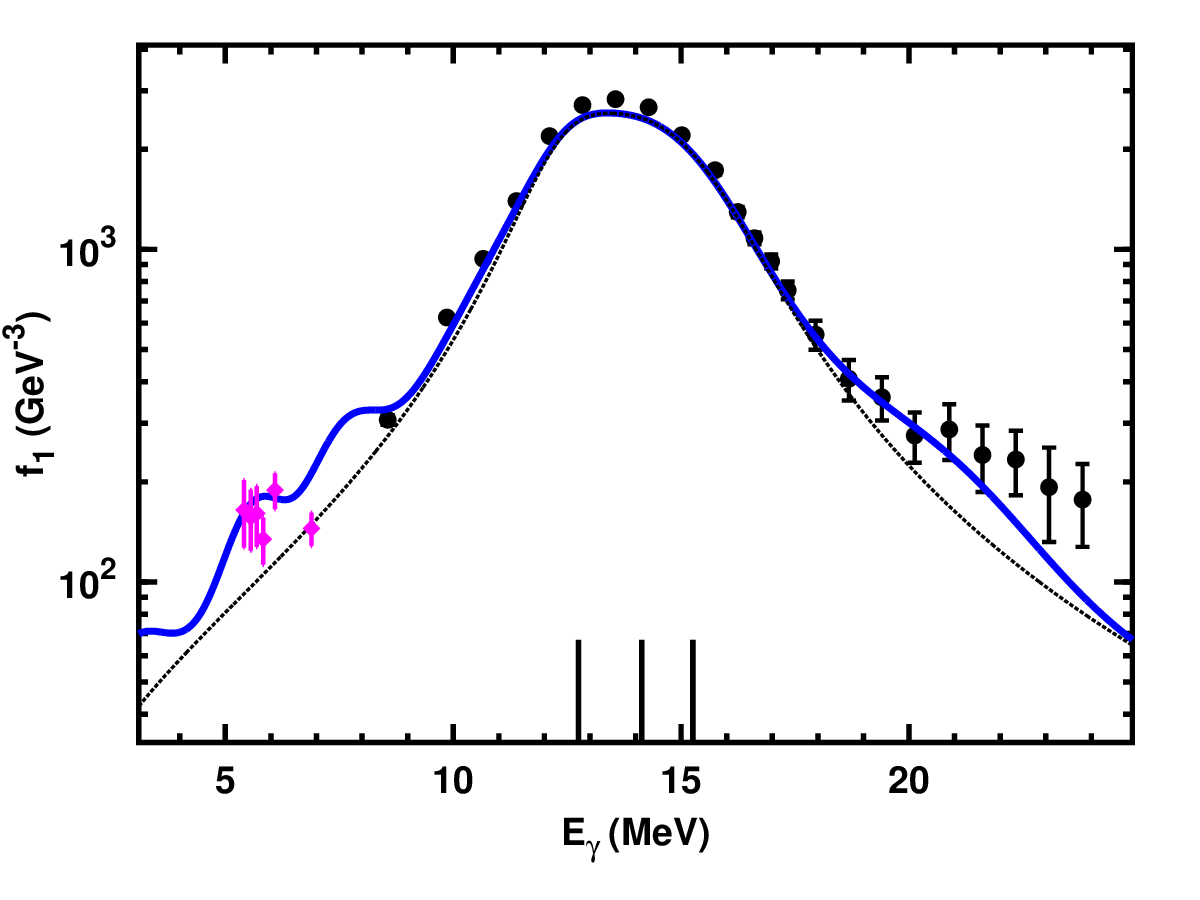}
\caption{(Color online) Photon strength for $^{190}$Os derived from ARC data (magenta diamonds \cite{ca79}) and from the cross section of photo-neutron production (black dots \cite{be79}). For comparison the TLO result is shown as before (see the caption of Fig.~\ref{figSe} ).  } 
\label{figOs}
 \end{figure}
 
\begin{figure}[ht]
\includegraphics[width=1\columnwidth]{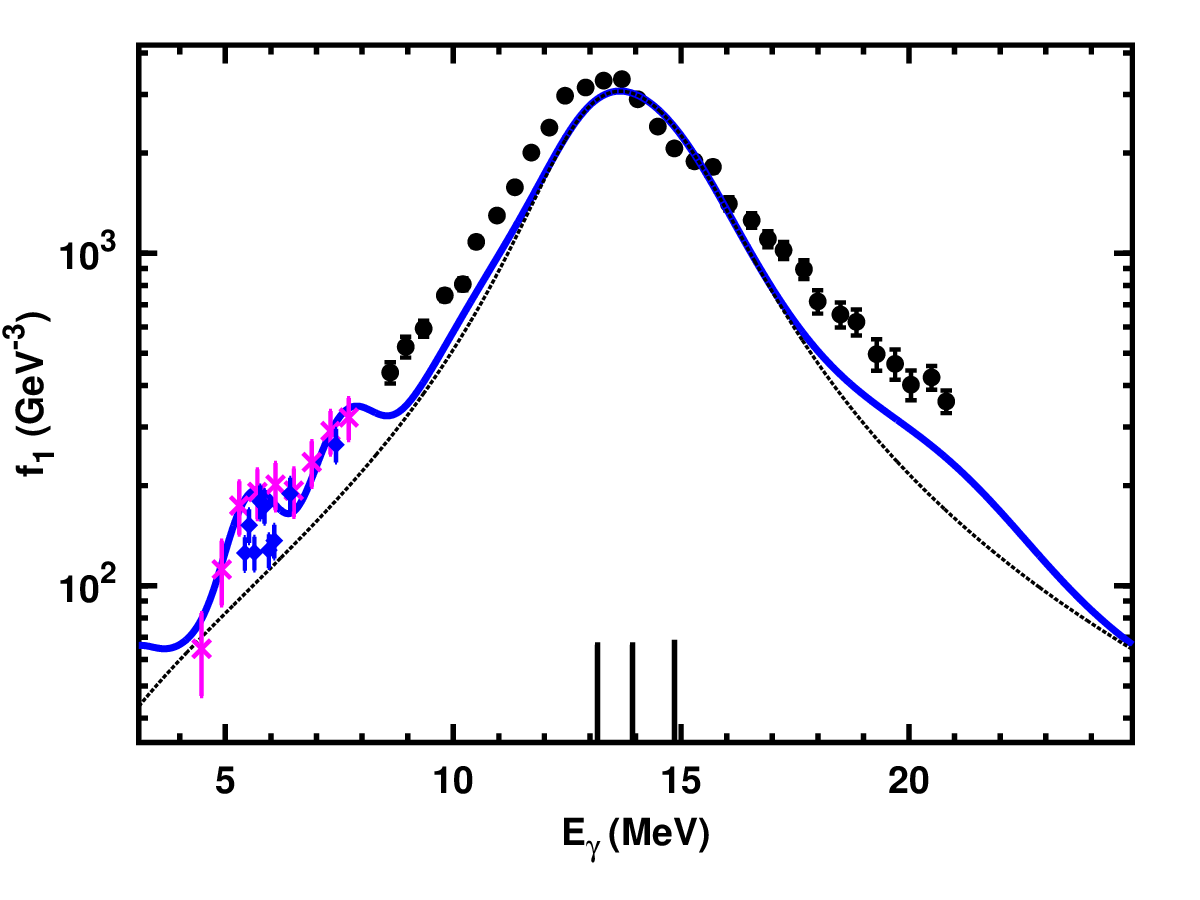}
\caption{(Color online) Photon strength for $^{196}$Pt derived from ARC data (\cite{mu00} blue diamonds) and from the photoneutron cross section (\cite{go78} black dots) in comparison to TLO (see the caption of Fig.~\ref{figSe} ). Photon scattering data (\cite{ma12} magenta x-symbols) are reduced by $30 \%$ in view of the "new" ansatz for the level density \cite{gr14}. } 
\label{figPt}
 \end{figure}

\begin{figure}[ht]
\includegraphics[width=1\columnwidth]{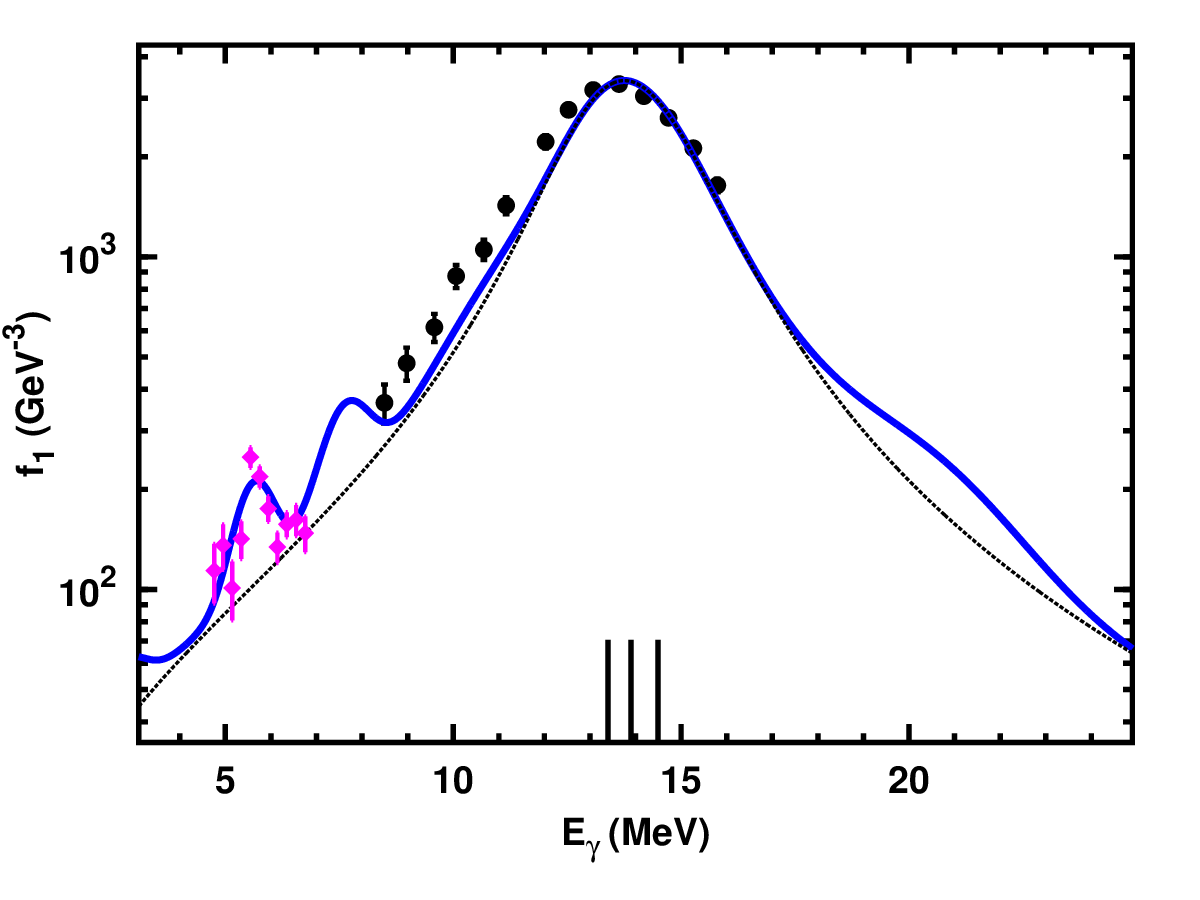}
\caption{(Color online) Photon strength in $^{202}$Hg calculated by TLO for the IVGDR (see the caption of Fig.~\ref{figSe} ) is compared to photon scattering (magenta diamonds \cite{la79}) and photoneutron production (black circles \cite{ve75, di88}), respectively, both studied with $^{\rm nat}$Hg. } 
\label{figHg}
 \end{figure}

As shown in  \cref{figOs,figPt,figHg} (and also in \cref{figTe,figEr}) photon scattering experiments covering the respective energies intermediate “pygmy” structures at $E_\gamma  \approx  0.4$ and/or $0.6  E_{IVGDR}$ are seen. It is noted here that enhanced gamma-strength was seen only at $E_x \approx 0.4  E_{IVGDR}$ from $\alpha$-scattering experiments with subsequent direct decay to the ground state \cite{po92, en09}, indicating an isoscalar character \cite{sa13} for this low energy pygmy mode - in accordance to recent calculations (\cite{re13}). 

Experiments with tagged photons \cite{la79} have identified a resonance-like structure in the cross section of photon scattering on targets in the vicinity of $^{208}$Pb, and the case of Hg (\cref{figHg}) is especially significant: TLO with its small $\Gamma$ predicts a considerably smaller cross section as compared to the low energy pygmy resonance near 5.6 MeV, which clearly surmounts the smooth tail. Here it was shown \cite{la79} to be important that photon scattering yields are properly corrected for inelastic scattering.

\begin{figure}[ht]
\includegraphics[width=1\columnwidth]{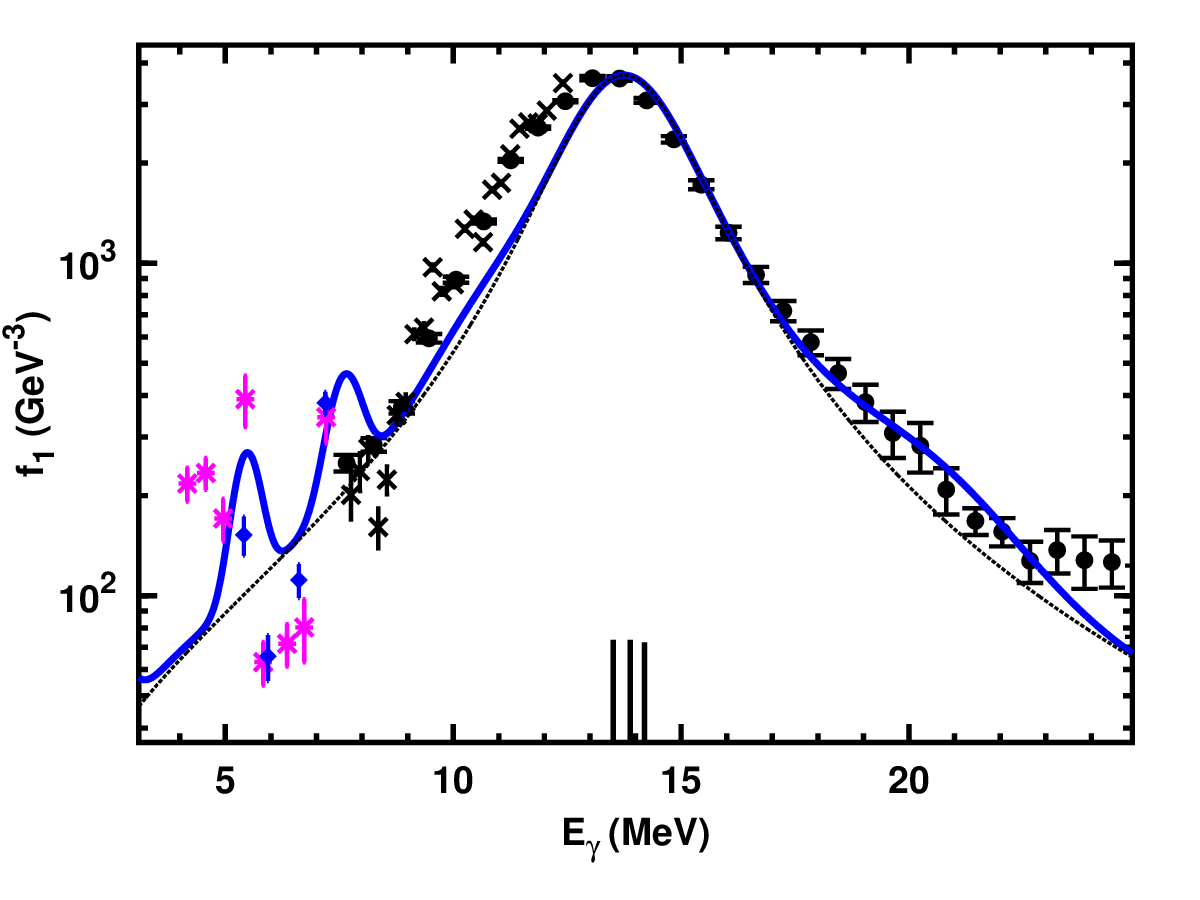}
\caption{(Color online) Photon strength for $^{208}$Pb derived from photon scattering data using a quasi-monochromatic beam \cite{la79}(blue diamonds), from bremsstrahlung \cite{sc12}(magenta star symbols) and from two photoneutron cross section measurements \cite{ve70} (black circles),  \cite{vy78} (black x). TLO for the IVGDR is depicted as described for Fig.~\ref{figSe} } 
\label{figPb}
 \end{figure}

In $^{208}$Pb (\cref{figPb}) the energy of such minor modes lies in a region of small level density and hence very large spacing between $1^-$-levels. Here p-h-excitations to $1^-$ may play a role as shown \cite{he14} to happen in $^{208}$Pb and it is intriguing to compare the strong $1^-$-level at 5.51 MeV \cite{pi09, sc10} to the strength function in the pygmy resonance in Hg: They are very similar in energy and energy-integrated strength, and can eventually be identified as one type of pygmy resonances \cite{zi05}.  In this mass region Porter-Thomas fluctuations have a large effect \cite{ve70, vy78, be82, na91} because of low level density reaching up to the IVGDR range, related to the large shell correction in near-magic nuclei; the averaging in our figures helps to see the smooth features. Indications of contributions from the GQR are observed as well and a peak near 20 MeV which we assign to the IVGQR was also indicated in inelastic proton scattering \cite{dj82}; unfortunately electron scattering data \cite{bu72, pi74, ku81} do not allow a fully consistent transfer of information. The ISGQR is indicated in \cref{figNd,figGd,figEr,figOs,figPt,figHg,figPb} to partly overlap the low energy slope of the IVGDR component with the smallest ${E_i}$. In nuclei with or near closed shells the calculations \cite{de10} predict more deformation as deduced within this study from the deformation induced splitting of the IVGDR. For closed shell nuclei a small deformation (reduced by a factor 0.4 to 1 as discussed together with Fig.~\ref{figSm}) results in a better description of the IVGDR peak shape; this reduction to $\beta=0.030$ leads to a value below the one derived from B(E2,${0^+ \rightarrow 2^+}$) \cite{ra01}. It disagrees to the popular belief that magic nuclei are spherical, but this has no significant influence on the strength in the tail region. 

\begin{figure}[ht]
\includegraphics[width=1\columnwidth]{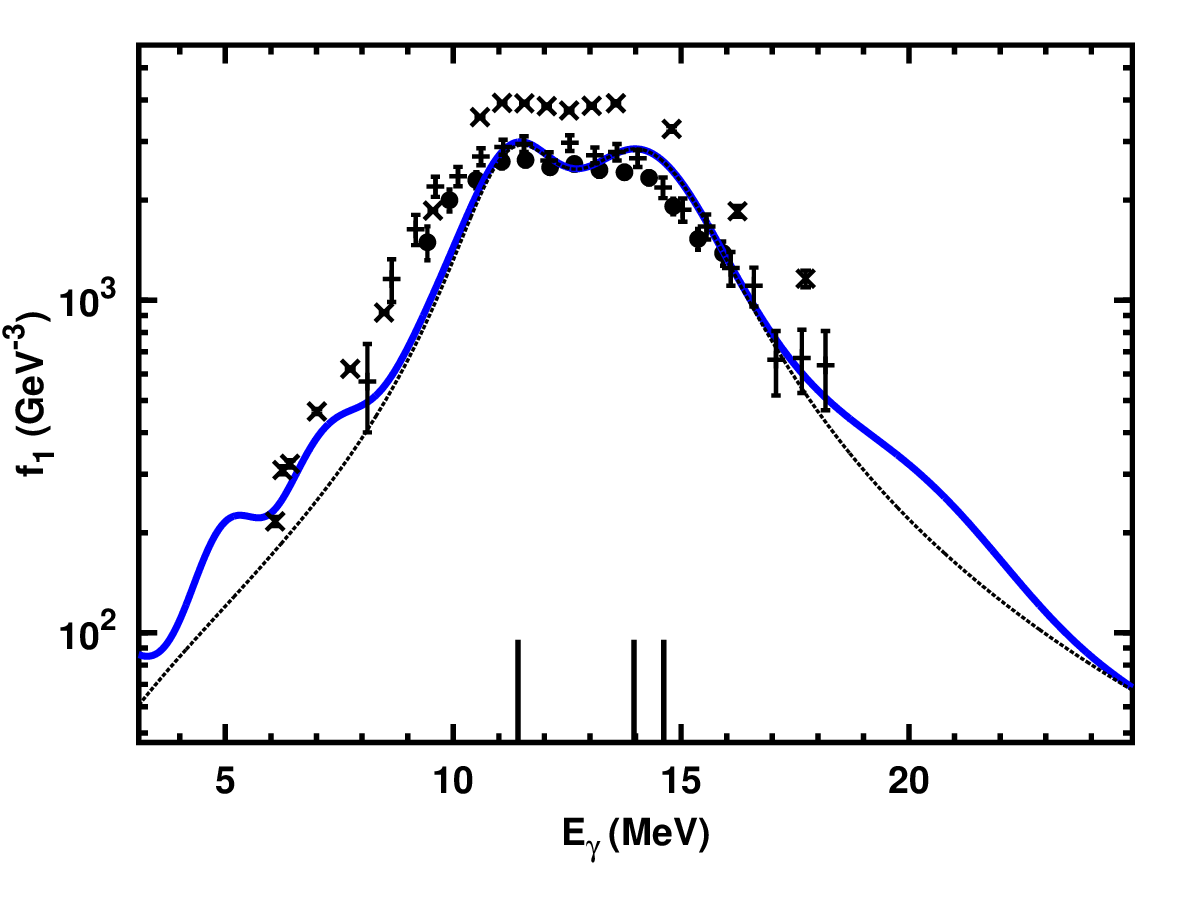}
\caption{(Color online) Photon strength for ${}^{232}$Th  derived from the sum of photoneutron and photofission cross sections from Saclay \cite{ve70} (black circles) as well as from Livermore \cite{ca80} (black x symbols). Photon absorption data (black $+$ symbols)
\cite{gu76} disagree to the latter, but agree to TLO (see the caption of Fig.~\ref{figSe} )}.  
\label{figTh}
 \end{figure}

\begin{figure}[ht]
\includegraphics[width=1\columnwidth]{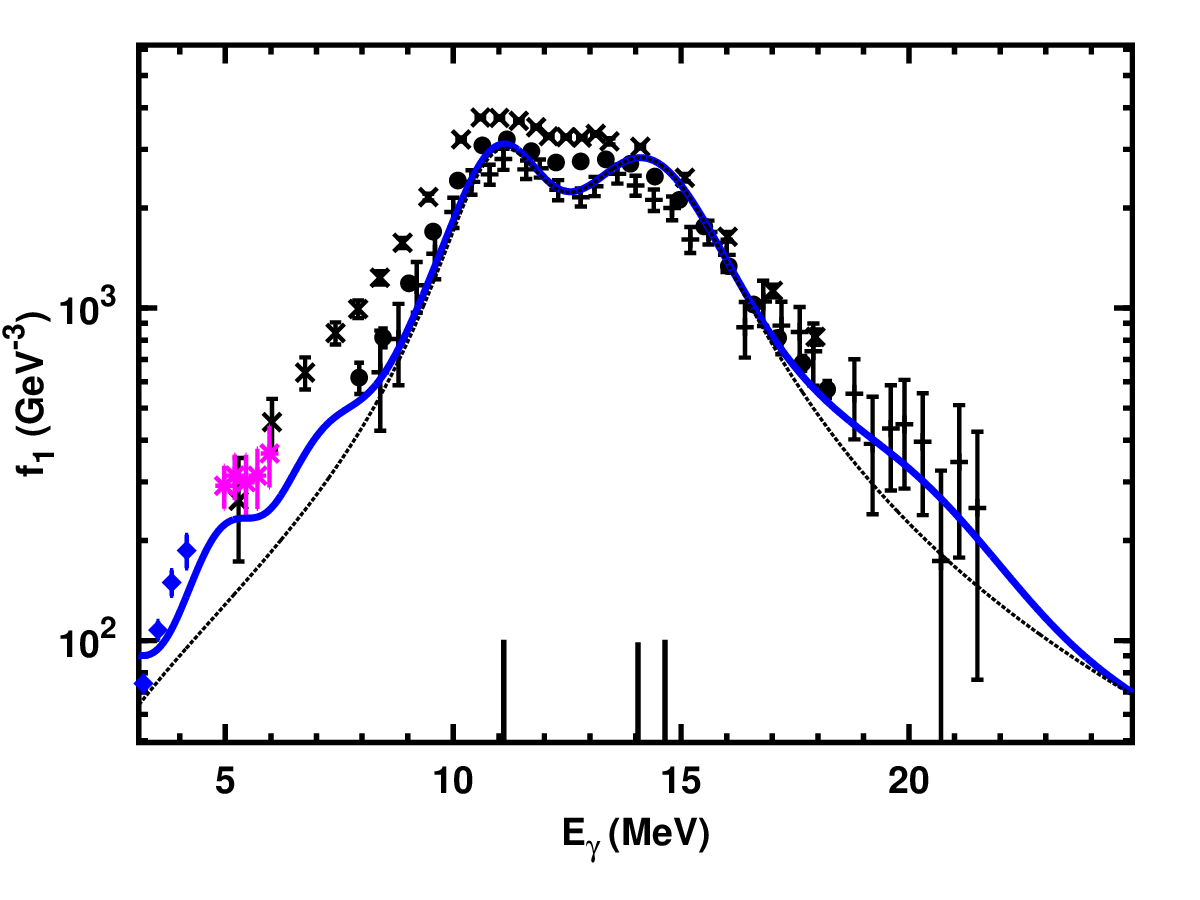}
\caption{(Color online) Photon strength for $^{238}$U predicted by from TLO for the IVGDR (see the caption of Fig.~\ref{figSe} ) in comparison to two differing sets of data derived from photo-neutron production (\cite{ve70} black dots )(\cite{ca80} black x symbols); both include photofission. Also shown are photon absorption results (\cite{gu76} black + symbols with large error bars) and data from photon scattering (\cite{bi87}magenta *-symbols) using a quasi-monochromatic beam. Data from $\gamma$-decay after deuteron-scattering \cite{gu14} at energies below 4 MeV are depicted as blue diamonds. } 
\label{figU}
 \end{figure}

In Figs.~\ref{figTh} and \ref{figU} three sets of experimental data for $^{232}$Th and $^{238}$U, respectively, in the range of the IVGDR are displayed together: The data with the large error bars were obtained by photon absorption. They agree within uncertainty to data stemming from a photo-neutron experiment performed at Saclay, which were reduced here by $10 \%$, as explained before. The agreement is not perfect, but indicates the reliability of both in the IVGDR regime.

The agreement between two data sets is important with respect to the disagreeing data obtained at Livermore \cite{ca80}. These cross sections for $^{232}$Th  and $^{238}$U are exceptional large in the sense, that an analysis on the basis of Eq. (\ref{eqIE1}) indicates an overshoot of $\approx 30 \%$ as compared to the TRK sum.

Low energy strength observed in actinide nuclei \cite{gu14} suffers from the missing parity assignment; a questionable choice for the IVGDR tail is likely to influence a comparison to previous data from e- and $\gamma$-scattering \cite{he88} assigning the multipole character as M1; some difference between these data is indicated.

\subsection*{Odd nuclei}
\label{subsubsec5}
\begin{figure}[ht]
\includegraphics[width=1\columnwidth]{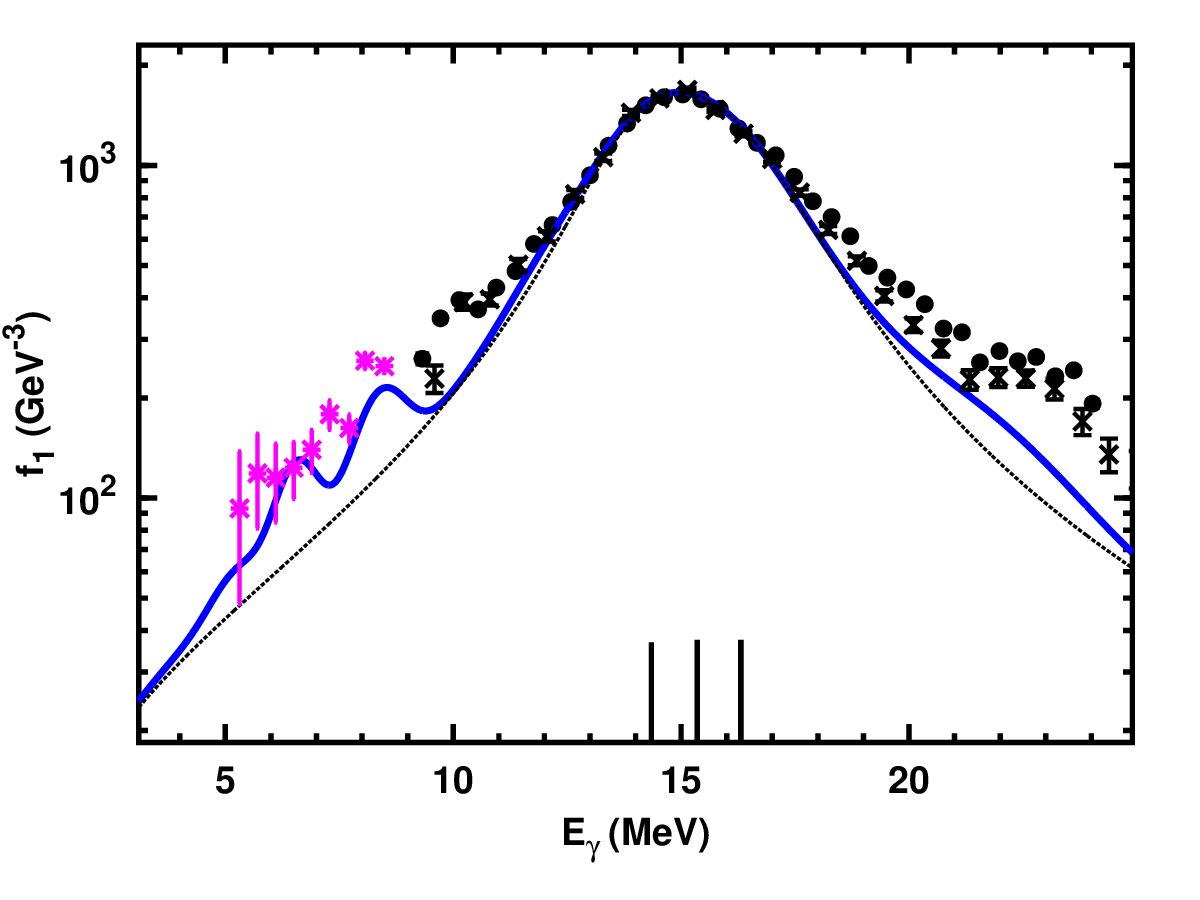}
\caption{(Color online)Photon strength for $^{127}$I calculated as described in the text (see the caption of Fig.~\ref{figSe} ), in comparison to two datasets of photoneutron production shown as black x-symbols \cite{be87} and black points \cite{be68}(reduced by $10\%$ like all data from Saclay; this factor reduces the disagreement to the other, newer measurement). The photo-absorption data below $S_n$ (magenta asterisk) are derived from elastic scattering by the neighbour nucleus $^{128}$Xe \cite{ma14}, E1 and M1 added and modified by 0.7.} 
\label{figIX}
 \end{figure}

\begin{figure}[ht]
\includegraphics[width=1\columnwidth]{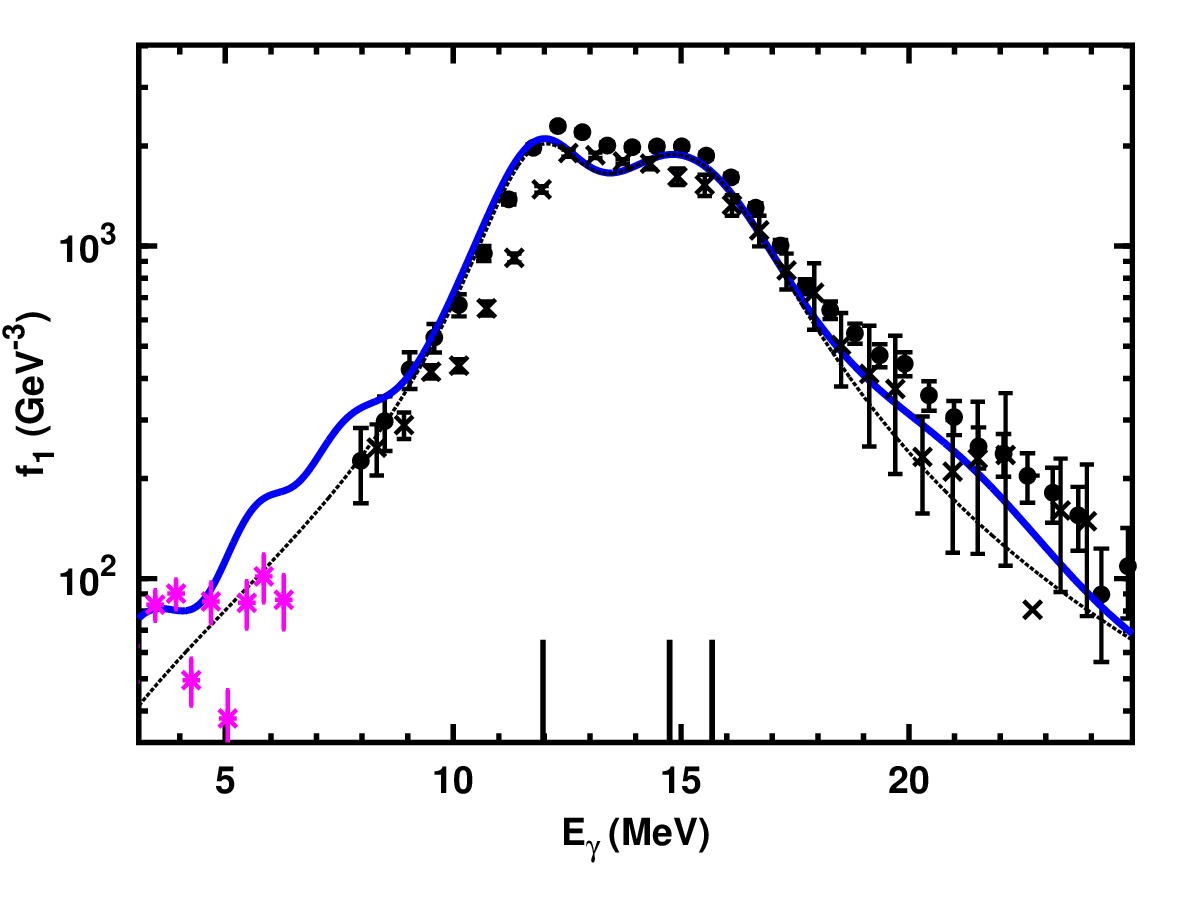}
\caption{(Color online) Photon strength derived for $^{181}$Ta from photon scattering \cite{ma14}(magenta *-symbols, below 7 MeV, modified by 0.7) and from photoneutron production (black x symbols, \cite{br63}; black dots, \cite{be68}) in comparison to TLO (see the caption of Fig.~\ref{figSe} ). } 
\label{figTa}
 \end{figure}

\begin{figure}[ht]
 \includegraphics[width=1\columnwidth]{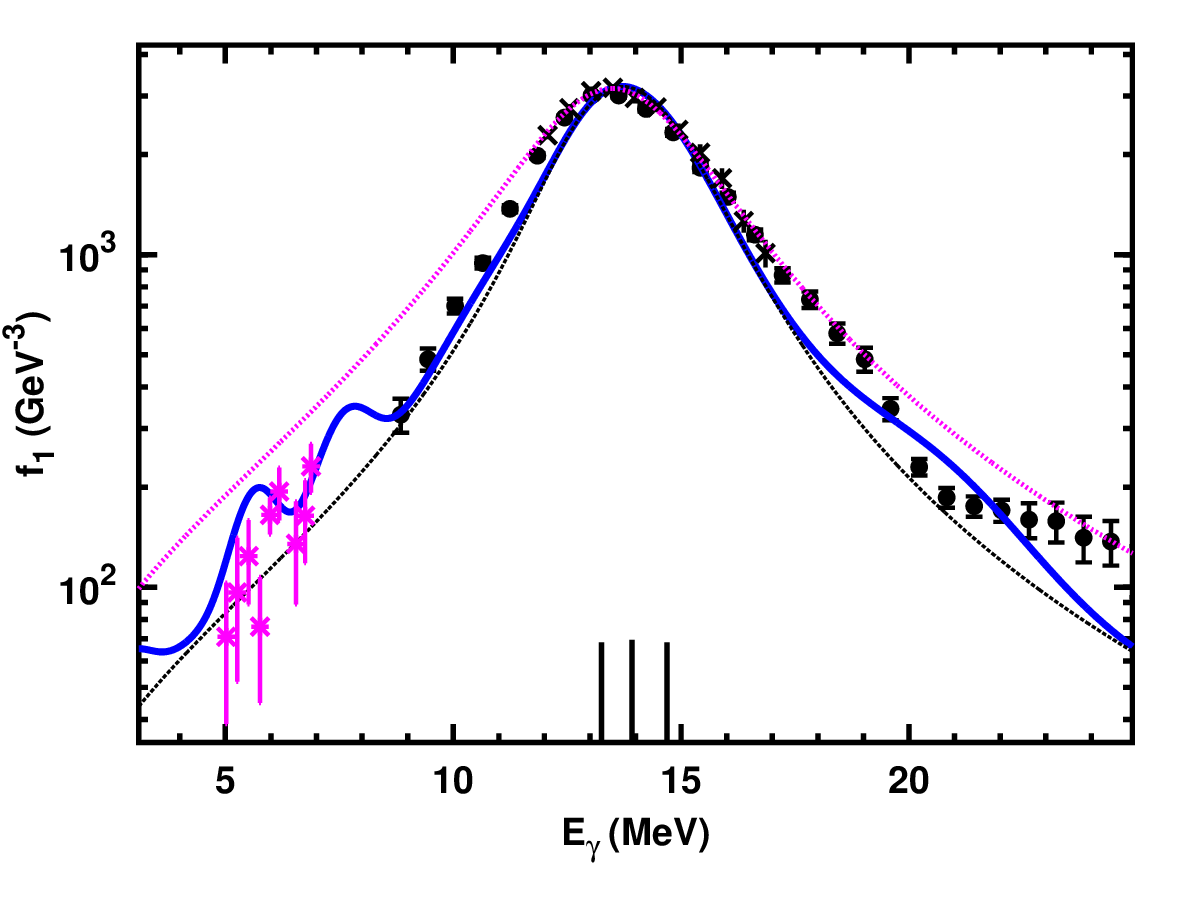}
\caption{(Color online) Photon strength for $^{197}$Au derived from photon scattering data \cite{ba73}(magenta asterisks) using a quasi-monochromatic beam and from the cross section of photoneutron production \cite{ve70}(black circles) in comparison to the TLO prediction (see the caption of Fig.~\ref{figSe} ). Also shown are newer data for the peak region \cite{be87}(black x- symbols) and a SLO fit curve from RIPL-3 \cite{pl11} (magenta dotted line; adjusted in height to experiment). Note that the close agreement between the two data sets is a consequence of the renormalization of the Saclay data. \cite{ve70}.} 
\label{figAu}
 \end{figure}

The CHFB calculations \cite{de10} on which our TLO parameterization is based have not yet been published for odd nuclei. As proposed \cite{ma16}, we use an average over the deformation parameters for even neighbor nuclei to obtain the oscillator frequency in Eqs. ((\ref{eqdIE}, \ref{eqEi} and \ref{eqflam})) to arrive at $f_{E1}$. It was also assumed, that for $J_0 \neq 0$ photon absorption into a mode $\lambda$ populates m members of a multiplet with m=min(2$\lambda$+1, $2J_0$+1) and the decay widths to the ground state $\Gamma_{0r}$ are equal for each member of the multiplet; the conditions for the validity of Eq. (\ref{eqflam}) are thus fulfilled. The strength observed corresponds to the cross section summed over the multiplet and this can be described by an effective $g$, which according to Eq. (\ref{eqGr0}) is:	
\begin{equation}
g_{eff}=\sum_{r=1,m} \frac{2 J_r + 1}{2 J_0 + 1} = 2\lambda +1
\label{eqgf}
\end{equation}
This ansatz is valid in heavy nuclei \cite{ba73} as it relates to the condition of weak coupling between the odd particle and the mode $\lambda$. The TLO-calculations for odd-$A$ nuclei as shown in \cref{figIX,figTa,figAu}
 were performed on the basis of Eqs.(\ref{eqdIE}, \ref{eqEi} and \ref{eqflam}) with $k=3$.	
 No extra spin dependent factors are needed and agreement to the experimental data is found to be similar as for even nuclei, also in the tail region below $S_n$. In Fig.~\ref{figIX} data for $^{127}$I are shown to be close to those for neighboring even nuclei depicted in Figs.~\ref{figBa} and \ref{figNd}. The agreement to TLO is obvious and also ``minor'' pygmy strength and the IVGQR are seen.
 For the nucleus $^{197}$Au not only the TLO-prediction is depicted, but also a SLO curve from RIPL-3 \cite{ca09, pl11}, which clearly over-predicts the data below $S_n$ extracted from Fig.~18 of ref. \cite{ba73}. In contrast to the missing strength as compared to a single Lorentzian (SLO) used there, the agreement is reasonable for TLO, as the discontinuity near 19 MeV may be related to the known  \cite{va14} incorrect separation of the 2n-channel.
 The inclusion of triaxiality in TLO leads to a reduction of $\Gamma_i$  and thus of $\sigma_{abs}$ for sufficiently large $(E_i^2-E_\gamma^2)^2$ in Eq. (\ref{eqdIE}). The  widths $\Gamma_i$  used previously \cite{ba73} are 2.9 and 4.0 MeV and an additional factor of 1.22 was obtained as compared to the TRK-sum rule. This factor is 1.0 for TLO and the values for $\Gamma_i$ are 2.7, 3.0 and 3.5 MeV. When $^{197,198}$Au are considered spherical
 $\Gamma \approx 4.5$ MeV results from a SLO-fit and this further increases the strength predicted in the tail region \cite{gu81}; similar conclusions follow from the use of the KMF-model for $^{197}$Au as was proposed \cite{ko90}. The satisfying agreement of TLO as presented in Fig.~\ref{figAu} favors our ansatz over the other models, especially when the strongly reduced number of fit parameters is regarded.
 \begin{figure}[ht]
 \includegraphics[width=1\columnwidth]{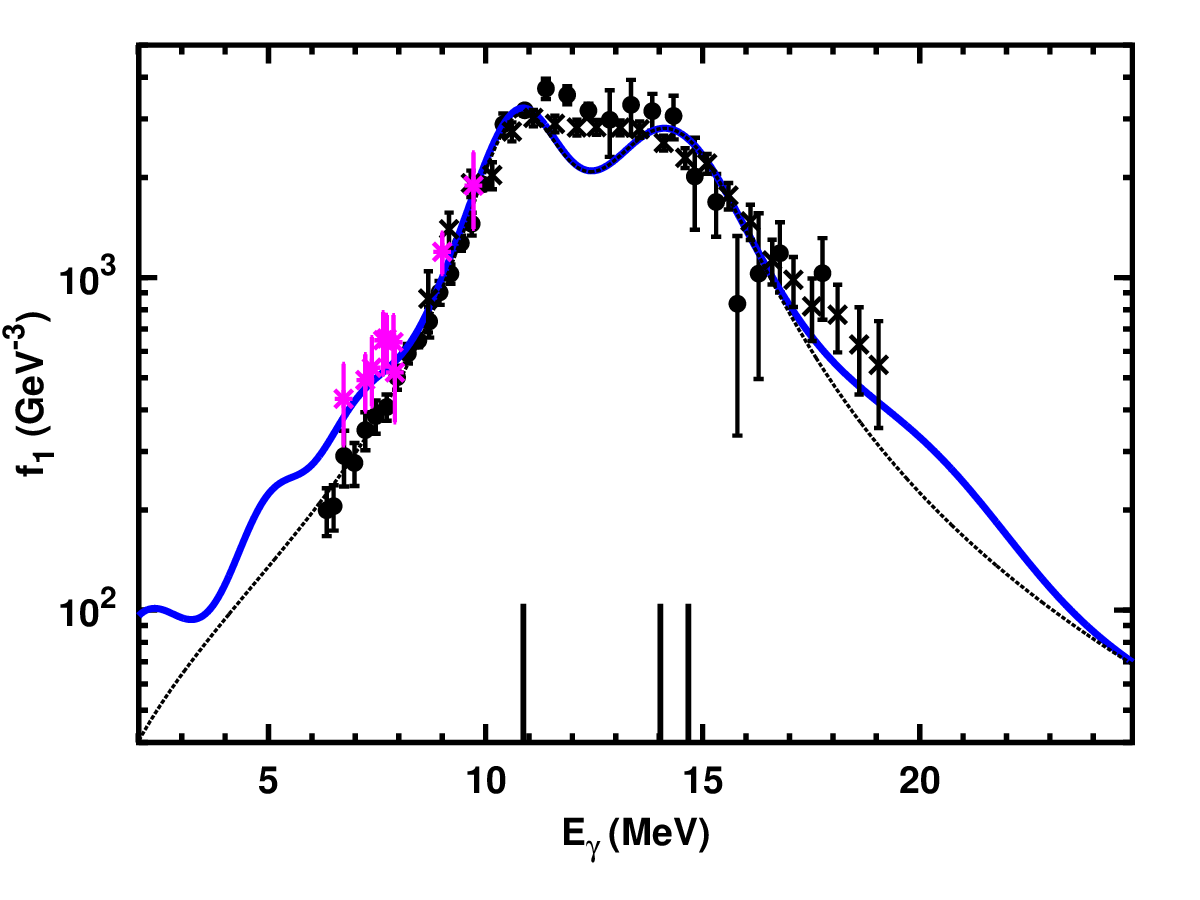}
\caption{(Color online) Photon strength for $^{239}$Pu (black dashed curve: TLO; full blue curve: minor strength added to TLO). Data are derived from summing cross sections of fission and neutron emission induced by quasi-mono-energetic photons \cite{be86} (black circles) and by discrete $\gamma$-rays from neutron capture \cite{am93}(magenta stars on the low energy slope). The absorption data from \cite{gu76} (black x) were obtained with bremsstrahlung. } 
\label{figPu}
 \end{figure}
 
In the bombardment with photons $^{239}$Pu mainly undergoes fission and the weaker neutron emission channel has to be added to obtain $\sigma_{abs}$. In Fig.~\ref{figPu} the result of this sum obtained in measurements at Livermore is compared to direct absorption data and a reasonable agreement is seen, as well as a good agreement to the TLO-prediction. This is remarkable in view of the disagreements depicted in Figs.~\ref{figTh} and \ref{figU} for the near neighbors $^{232}$Th and $^{238}$U and doubts about these older data seem justified. 
Together with the agreement between $^{88}$Sr and $^{89}$Y reported recently \cite{gr14} the examples presented in this section support the TLO ansatz for the derivation of photon strength in odd nuclei. Hence for all nuclei presented here a quite good agreement between observations and the TLO prediction for E1 strength is found, if minor strength is accounted for at least approximately – for which we present a phenomenological solution. A description of the IVGDR in deformed nuclei \cite{ma05} not using the TRK sum rule and not based on a fully self-consistent calculation of the shape parameters was by far less successful in its predictions.

\section{Summary}
\label{sec7}

The results of the comparison of experimental data for more than 20 nuclei in the mass number range from 78 to 239 to Eqs.(\ref{eqdIE} and \ref{eqflam}) can be summarized as follows:
\begin{enumerate}[(a)]
\item The centroid IVGDR energies derived from droplet model fits to masses \cite{my77, mo95} are in accord to the data, when an effective mass $m_{eff}c^2 = 800$ MeV and the proton radii as predicted from the CHFB calculations \cite{de10} are used.
\item There is no indication of a strong departure from the classical dipole sum rule \cite{ku25, re25, ge54} and the yield at energies above the IVGDR is assigned to the quasi-deuteron strength. Good conformance with the TRK-sum is not only observed near closed shells or for nuclei with large $|Q_0|$. Previously reported  \cite{be75, di88, ut11} local fits together with the {\em ad hoc} assumption of spherical or axial symmetry overestimated IVGDR widths, resulting in the excess over TRK; this indicates the importance of extra information on shapes \cite{ju08} and especially the broken axiality as predicted by CHFB-GCM calculations \cite{de10}. 
\item The resonance widths vary only smoothly with $A$ and $Z$ as predicted by hydro-dynamical considerations \cite{da64, na91}. Only by allowing broken axial symmetry two experimental observations are described well: (1) three rather narrow Lorentzians add up to an IVGDR structure with a width as large as apparent and (2) for all nuclides regarded the spreading widths of the three components only depend on the resonance energies via a power law, ({\em  cf.}  Eq. (\ref{eqEi})), with an exponent of 1.6, as predicted for triaxial shapes \cite{bu91}.
\item Having $\Gamma$ depend on $E_i$ only, (\ref{eqEi}) causes the two upper resonance parts (and their sum) in nuclei with large $Q_0$ (see \cref{figGd,figEr,figTa,figPu,figTh,figU} to have reduced height albeit all three components have equal strength. The IVGDR data together with data for high as well as for lower energies do not allow for a strong variation of the width with photon energy, as previously postulated \cite{ka83}.
\item The Axel-Brink hypothesis and the TRK sum rule together with the photon-energy independence of the IVGDR width are essential for our TLO-ansatz, and no clear hints for deviations are seen when comparing data to TLO. In $\gamma$-decay spectra observed after averaged resonance neutron capture (ARC) as well as in photon scattering two E1 pygmy modes are seen in nuclei with $A> 70$ (if the respective energy range is covered); both kind of data are in reasonable accord to each other.
\item If an enhancement by intermediate structures is handled as minor strength similar as in previous work for nearly magic nuclei \cite{ax70,la79} also the lower energy dipole strength in nuclei of intermediate deformation (\cref{figNd,figBa,figIX,figOs,figPt,figAu}) is reasonably well accounted for in TLO (plus 'minor' strength, cf. Table~\ref{tab1}) – with a much lower number of parameters in comparison to previous work using local fits to data\cite{ko90, ko93, mu00}. 
\item  For lower energies the scissors M1 mode was predicted \cite{he10} to also be a general feature in heavy nuclei, but in photon scattering a clear separation from the E1 strength originating from quadrupole-octupole coupling is needed. At such low energy $\alpha$-cluster excitations are predicted \cite{sp15} to show up as well. 
\item The higher pygmy photon strength may well be due to an isoscalar vortical proton motion \cite{ry02, re13}. A vibration of excess neutrons against a core, was predicted \cite{mo71} to appear below $\approx 0.4  E_{IVGR}$ and to cause excess above the Lorentzian tail; but in this energy range E1 strength may as well be related to neutron p-h modes. 
\item  Exact deformation parameters have stronger influence on the peak of the IVGDR than on its tail, but the width influences the photon strength and TLO uses photon-energy independent and small damping widths depending on the resonance energies $E_i$ only. This is an important fact for the prediction of neutron capture cross sections and respective data are well reproduced by giving up the assumption of axiality  \cite{gr14}. Experimental data below $S_n$ used as an argument for alternative parameterizations \cite{ko90, mu98} can equally well be described by TLO, if the extra 'minor' strength is added (cf. Figs.~, \ref{figGd} to \ref{figPt}).
\item  An account for the variance of the deformation parameters by instantaneous shape sampling \cite{zh09} only leads to small changes of the calculated strength in the IVGDR for deformed nuclei: The resonances are widened near the peak region, but the low energy tail remains unchanged.
\item At variance to circumstantial evidence often quoted as proving axiality for most heavy nuclei, this paper together with our previous work on nuclear level densities \cite{gr14} indicate that the assumption of an axial shape is not a good approximation for very many heavy nuclei even in the valley of stability. This faalsification of a common praxis resembles to what was predicted for broken symmetry in crystalline matter by Jahn and Teller in 1937 \cite{ja37}, only much later observed and demonstrated to also apply to sphericity and axiality of nuclei \cite{re84}.
\end{enumerate}

A very similar approach with a comparison to data for a number of isotopes was published previously \cite{ju08, be11, na08, er10, sc12, na10, ju10, gr11, gr17}; these papers are like a part of the present work, such that an impressively wide sample of IVGDR data in heavy nuclei is shown to be well described by TLO. 

\section{Comparison to other work}
\label{sec8}

We consider our phenomenological approach an intermediate step  between microscopic theory and the 'classical' analysis by Lorentzian fits \cite{be75, bo75, di88}. Based on the three-dimensional nature of nuclei and inspired by Coulomb excitation results \cite{cl86} our TLO ansatz seeks for a description with three pole energies for all heavy nuclei; this is in accord to a microscopic calculation, which is available for many nuclei and we take the deformation parameters from it. We had to extend our macroscopic approach that way, as symmetry breaking is likely to be induced by quantum effects \cite{ja37, re84} like nuclear shells, and these are not contained in macroscopic liquid drop (LDM)-schemes. We note that LDM predictions of ground state masses have only little sensitivity to deformation values \cite{mo08, ra01}. Thus a comparison of the TLO scheme to other work has to regard both, macroscopic as well as microscopic work.

\subsection*{Analysis by one or two Lorentzians}

\begin{figure}[ht]
\includegraphics[width=0.95\columnwidth]{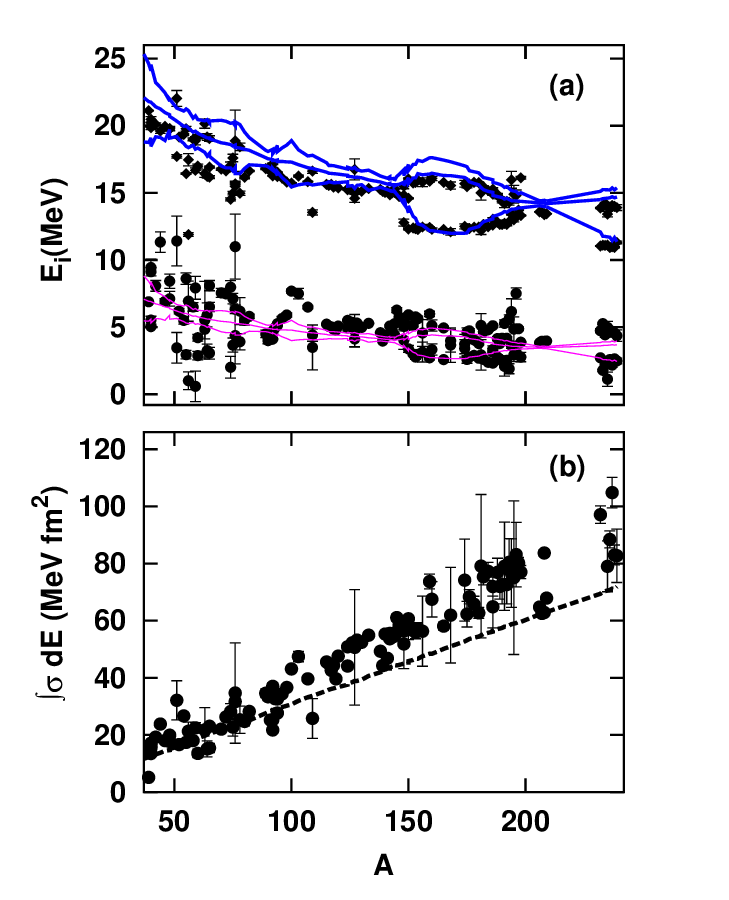}
\caption{(Color online) Panel (a) shows the energies (top) and widths (bottom, both vs. mass number) resulting from $\chi^2$-fits to the IVGDR in heavy nuclei, as  compiled recently \cite{pl11}. The fits are based on one or two Lorentzians and two points per nucleus are shown, if a two-pole fit led to a smaller  $\chi^2$. Our calculations with three poles (TLO) as described in section~\ref{sec4} are depicted as three drawn curves for the three components of the IVGDR's; (blue for $E_i$ and magenta for $\Gamma_i$). In panel (b) the resulting GDR-integrals as obtained by the Lorentzian fits from ref. \cite{ca09} are depicted in comparison to the TRK sum rule (dotted line), which is surpassed considerably in most cases, whereas TLO obeys it by definition. } 
\label{figPlu}
 \end{figure}

Starting from LDM concepts IVGDR's in heavy nuclei have been analyzed repeatedly by assumig either spherical or axial symmetry, {\em i.e.} with a single Lorentzian or a sum of two \cite{be75, ca74, be79, ju08}. Recently a new analyis along these lines was performed and reviewed within the RIPL-3 project \cite{ca09, pl11}. Results from local fits limited to the region near the IVGDR peak  are shown in Fig.~\ref{figPlu}  which are at variance to this work and  earlier publications by us. \cite{ju08, er10, ju10}. 
The upper panel of Fig.~\ref{figPlu} shows how the energies and Lorentz widths from the local fits as published within RIPL-3 \cite{ca09, pl11} scatter as compared to the description of the IVGDR shapes, allowing three poles and thus a ‘triple’ Lorentzian (TLO) parameterization, derived from the global fit procedure detailed in section~\ref{sec4}. As seen from Fig.~\ref{figPlu} the TLO-method results in a smooth dependence on $A$ which is modulated only due to variations in shape, as presented in \cref{sec4,sec6}. As shown in the bottom part of Fig.~\ref{figPlu}, the TRK sum rule, Eq. (\ref{eqIE1}), disagrees in many nuclei to the Lorentzian fits \cite{di88, pl11} performed for the data of each nucleus independently without account for the possibility of broken axial symmetry. In these fits the width parameter was adjusted for each isotope separately to fit the peak region and for $A$ between 90 and 150 a  large discrepancy is observed as well as wide fluctuation with $Z$ and $A$ of this apparent width indicating a non-systematic variation which is difficult to conceive within the spreading concept. A similarly erratic dependence of the integrated IVGDR strength on $Z$ and $A$ was reported \cite{pl11} to result from this approach of fitting the photo-absorption data locally. In some cases the integrated cross section overshoots the classical sum rule given by Eq. (\ref{eqIE1}, first term) by up to $50\%$. Apparently the two problems are closely related, as the resonance integral is proportional to the product of height and width. The large unsystematic scatter obtained in these local fits speaks against their use {\em  e.g.} for nuclear astrophysics, but it has been proposed \cite{gy98} to be used in that field. We point out, that admitting the breaking of axial symmetry, albeit often weak, has a clear advantage here as a triple Lorentzian (TLO) improves a description of IVGDR-shapes for those heavy nuclei for which data exist. And it is easily extended to others, as the contribution to it, which requires deformation values as derived microscopically are widely available \cite{de10}. They determine the three resonance pole energies and they also enter our prediction for the IVGDR widths: the energy dependence derived from regarding its value for different $A$ and $Z$ was found to be well represented by a dissipation model \cite{bu91} and we transposed this dependence to a variation between the three Lorentzians in one nucleus. The factor $c_w$ in Eq. (\ref{eqEi}), quantifying the width of each Lorentzian, was first used by us as a quantity to be fitted.   

\subsection*{Microscopic descriptions of the IVGDR}

Later we realized the good agreement to the spreading width obtained by an optimisation of the optical potential in a microscopic calculation for $^{208}$Pb \cite{do72}. From this we conclude a good agreement of our TLO to this special RPA (random-phase approximation) calculation in a case not hampered by significant deformation effects. In heavy nuclei in general the apparent irregular A-dependence of the width seems to have motivated modifications to RPA based studies of the damping in collective nuclear dipole vibrations \cite{be83}. But TLO puts this into a different perspective: The irregularity is strongly reduced for all heavy nuclei when making a global fit using three Lorentzians.
Keeping this in mind we now regard a few published examples for RPA-predictions on IVGDR strength, energy and spread of width. We find that most of them an integrated strength in accordance to the TRK sum rule was found. As discussed in section~\ref{sec3} an accuracy below $30 \%$ requires a knowledge about ph-correlations in quasi-deuteron pairs, whereas pygmy structures in the low energy tail contribute about a factor of 10 less. \\

Microscopic results on IVGDR energies and widths relate to one- and two-body dissipation and for both results from the theory of Fermi liquids may be applied for nuclei. In an early relevant study \cite{fi86} these two components are documented to both have their origin in work of Landau; this is why we avoid to connect one of them with his name. In the list of origins for the IVGDR widening in section~\ref{sec4} we list these two contributions under (c) and (a), respectively; for the one body term we use the previously proposed \cite{be10} expression fragmentation. Under (b) we have a separate component for nuclear shape induced splitting, which when added to the others may eventually be confused with fragmentation caused by shell effects. Work to identify an IVGDR fragmentation in $^{208}$Pb has been performed experimentally\cite{ve70, ca74, vy78} as well as by an advanced shell model calculation \cite{br00, sc10}. As shown in fig. 5 of ref.\cite{sc10} it did not indicate a  significant fragmentation, whereas the QRPA study also shown compares less well to the shell model calculation, and thus is at variance to a schematic Tamm-Dankoff prediction \cite{be11}. This has led to our assumption for TLO to consider the fragmentation, which has been labelled \cite{be83} Landau damping, to be less important than the shape induced splitting. 
This as well as the treatment of 2p-2h and higher particle-hole states constitute the main difference of TLO to RPA work: In many RPA publications an
additional width of up to 2 MeV has been folded to the predicted strength distribution to improve the agreement to experimental data. In TLO the spreading width is transferred from a calculation for $^{208}$Pb ($\approx 3$ MeV) \cite{do72} by using Eq.(\ref{eqEi}); for the nuclei studied by TLO it is considerably larger than 2 MeV, and a good match to experimental data is observed.
And in both the escape width is neglected, as previously proposed as well for heavy nuclei \cite{be83, wo91}. As the deformation parameters used for TLO stem from a CHFB+GCM calculation a future merge of RPA calculations and our ansatz seems quite possible. It may start from a microscopic derivation of the fit parameters in Eq.(\ref{eqEi}) extracted globally. We believe that the the role of broken axial symmetry should be regarded in detail as the simple fits to only one or two IVGDR Lorentzians result in an apparent width without any clear $N$ or $A$ dependence \cite{pl11}. The surprisingly large fragmentation reported for more or less all the microscopic calculations has to be understood better: Is it due to a cut-off in the number of shells included or is the combination of calculational approximations - like axiality and RPA - responsible? \\

In addition to the QRPA for $^{208}$Pb mentioned above there is a number of other RPA calculations published for this nucleus, but their discussion would exceed the scope of this paper. Instead we now regard the agreement of similar and rather new calculations to data for nuclei away from magic shells, as we have mainly studied those with TLO. Recently investigations \cite{ma16} in the framework of axially symmetric deformation and the finite-range Gogny force were performed for about 30 nuclei with $70<A<240$. They result in discrete E1-strength distributions which are usually folded by a Lorentzian with 2 MeV width and shifted by an energy of up to 3 MeV - both adjusted arbitrarily  to arrive at a reasonable agreement to the data. With this quasiparticle RPA (QRPA) approach based on an axially symmetric-HFB equation the influence of various calculational parameters was studied. It was found that the number of harmonic oszillator shells and the upper energy cut-off as well as the choice of the Gogny parameters influence the resulting shapes and the predicted strength in the IVGDR considerably. Three different approaches to theoretically predict the energy shift were tested, but for none of them a clear preference was expressed and also none of the two choices for the Gogny force was favoured clearly, albeit one of them results in a better fit to ground state masses. This publication \cite{ma16} itself as well as regarding the differing results of others give a good impression on the uncertainties to be reduced in future work. \\

One other calculation \cite{kl08} to be mentioned uses the zero-range Skyrme force SLy6 as approximation for the nucleon-nucleon interaction and makes predictions for the IVGDR's in about 20 well deformed nuclei with $A$ ranging from 156 to 238 and use is made of an earlier optimisation for the 'best' force to predict IVGDR energies within 1 MeV. To evaluate the sum rule the work makes a fit to the data which obviously disregards the contribution from the quasi-deuteron effect discussed in section~\ref{sec3}. Resulting discrete line spectra are averaged with $\Delta=$ 0.5, 1 or 2 MeV using Lorentz functions and compared to data. A reasonable agreement is only obtained using 2 MeV for $\Delta$, which the authors distinguish from the two resonance widths $\Gamma_1$ and $\Gamma_2$ that appear in their approach assuming axiality. The large difference between the two and the meager agreement to data for less deformed nuclei apparently indicate the effect of disregarding triaxiality. Here it should be repeated that triaxiality is enlarged in HFB-calculations \cite{ha84} with spin projection before the variation, and this finding was recently put into perspective by a covariant density functional approach \cite{bo16}. \\

The aspect of triaxiality was also regarded \cite{be10} and related to IVGDR's in about 20 not strongly deformed nuclei by introducing values for $\beta$ and $\gamma$ from an IBA-analysis of spectroscopic data into a QRPA calculation; an ISS treatment similar to the one described in section~\ref{sec4} was applied. The calculations result in a considerable fragmentation and a comparison to experimental IVGDR data was made only for 5 Sm-isotopes after a folding with an energy dependent Lorentzian width of approximately 2.5 MeV, apparently selected arbitrarily; the agreement is difficult to judge because of a small linear scale, but it seems reasonable.  

\subsection*{Low energy tail}

A straightforward prediction of the IVGDR spreading width into the the low energy tail becomes possible for any nuclide through our explict acount for triaxiality \cite{gr14}. We thus extrapolated the electric dipole strength to lower energy where it is important for radiative neutron capture \cite{gr14, ju10, ju17}. We showed \cite{sc12, gr17} the influence on this process to be centered below $E_\gamma= 4 $ MeV, where the intensity of primary photons is predicted to peak. At variance, previous conclusions are made based on information \cite{ca09, ma16} (cf. figs 41 and 14, respectively) not distinguishing between primary and secondary transitions and not caring about effects of the detection limit. For Zr-isotopes \cite{ut08} information was mentioned on a rather low E1-component as derived from a HFB+QRPA calculation \cite{gy04}, proposed \cite{ca09} by RIPL-3; the resulting suitable agreement to capture data apparently needed a surprisingly high magnetic strength, which superseeds respective systematics \cite{he10} by a factor of $\approx 3$ and hence the underlying E1 model may be questioned. Apparently any prediction on E1-strength in the IVGDR-tail has to be completed by a derivation of strength of other nature. The abovementioned calculation \cite{be14} involving IBA discusses low energy strength in detail, but presents data only above 7 MeV. For four isotopes of Sn predictions were made recently in an relativistically covariant approach \cite{li09} which covers photon energies down to zero, but no comparison to experiment is presented; the reported photon strength near 4 MeV (important for radiative capture of s-process neutrons) agrees to our systematics in the valley of stability better than earlier predictions. For the nucleus $^{206}$Pb a detailed study compares new data obtained with laser backscattering photons to an energy-density functional plus quasiparticle phonon model \cite{to17}. The experimental strength we report for the lead region (cf. Figs.~, \ref{figPt} to \ref{figPb}) compares better to the clustering of peaks in the experimental spectrum than to the one in the calculations; unfortunately a detection limit does not allow to report a strength function in absolute units. Experimentally the situation at low energy is hampered by the rarity of absolute strength values in and around pygmy structures, for which very often only energy information is available from experiments \cite{sa13}. To arrive at the phenomenological description as proposed by us we had to regard rather old work and this scan produced the respective data points in our figures. We hope that our compilation of information on an absolute scale may inspire respective microscopic calculations.

\section{Conclusions}
\label{sec9}

In our final conclusion we stress three points connected to broken axial symmetry:
\begin{enumerate}
\item Replacing {\em ad hoc} assumptions on nuclear deformation by shape information from independent sources the description of IVGDR observables can be improved considerably: Combining the theoretical prediction of triaxiality in all heavy nuclei as derived from a CHFB/GCM calculation \cite{de10} to the power law for non-axial shapes \cite{bu91} results in good resonance energies and widths without adjusting parameters locally.  \\

\item The triple Lorentzian (TLO) fit to IVGDR`s is global as it has only two free parameters – one each for energy and width, both depending on A and Z only indirectly via $E_i$ – and it still allows the TRK sum rule to be fulfilled. We stress that our good representation of IVGDR data for nuclei with intermediate deformation is at variance with the often made assumption of axiality. Three or more local parameters are required to obtain a good fit for each one nucleus \cite{ca09, pl11} under the axial approximation; thus we propose to validate it carefully before applying it. \\

\item  Using TLO with the IVGDR width not modified by an extra photon energy dependence leads to agreement with experimental data. The application of the KMF-model prediction \cite{ka83} to the IVGDR in such a way that its width becomes directly dependent on the photon energy is shown in Fig.~\ref{figPh} to falsely cause a too high tail above it. We have demonstrated alreadly previously \cite{gr14, ju17, gr17} how a triple Lorentzian (TLO) without such a width-variation and the respect of axial symmetry breaking improve the prediction of neutron capture yields of importance in nuclear astrophysics. 
\end{enumerate}

\section*{Acknowledgements}
This work is supported by the German federal ministry for education and research BMBF (02NUK13A) and by the European Commission within the 7th EU framework programme under ERINDA (FP7-269499) and through Fission-2013-CHANDA (project no. 605203). Intense discussions within these projects and with other colleagues, especially with Ronald Schwengner, Julian Srebnry and Jonathan Wilson, are gratefully acknowledged.

%

\input{breakref171203.tex}
\end{document}

%% file: breakref171203.tex
\newpage{\pagestyle{empty}\cleardoublepage}